\documentclass[fleqn,usenatbib]{mnras}

\usepackage{newtxtext,newtxmath}

\usepackage[T1]{fontenc}
\usepackage{ae,aecompl}

\usepackage{graphicx}	
\usepackage{amsmath}	
\usepackage{amssymb}	
\usepackage{longtable}
\usepackage{rotating}
\usepackage{subfig}

\newcommand{\nii}{[N II]}
\newcommand{\ha}{H$\alpha$}
\newcommand{\logniiha}{log$_{10}$(\nii/\ha)}

\title[SparsePak Survey of Nearby Galaxy Mergers]
{Dynamics and Shocks from \ha\ Emission of Nearby Galaxy Mergers}

\author[S. A. Mortazavi et al.]{
S. Alireza Mortazavi,$^1$\thanks{s.alireza.mortazavi@gmail.com}
Jennifer M. Lotz,$^{2,1}$
\\
$^{1}$Department of Physics and Astronomy, Johns Hopkins University, 3400 N Charles St., Baltimore MD, USA\\
$^{2}$Space Telescope Science Institute, 3700 San Martin Dr., Baltimore MD, USA}

\date{Accepted XXX. Received YYY; in original form ZZZ}

\pubyear{2017}

\hypersetup{draft}
\begin{document}
\label{firstpage}
\pagerange{\pageref{firstpage}--\pageref{lastpage}}
\maketitle

\begin{abstract}
We examine the dynamical properties of interacting galaxies and the properties of shocked gas produced as a result of the interaction. We observed 22 galaxy mergers using the SparsePak IFU at Kitt Peak National Observatory (KPNO). The goal of the observations was to obtain the \ha\ velocity maps over the entire luminous parts of the galaxies including the faint tidal tails and to find extended shocks and outflows. Our sample consists of major and minor galaxy mergers with mass ratios $1<\mu<8$. We fit multiple kinematic components to the \ha\ and \nii\ emission lines, develop an MCMC code to robustly estimate the error of fit parameters, and use the F-test to determine the best number of kinematic components for each fiber. We use \nii/\ha\ and velocity dispersion of components to separate star-forming (HII) regions from shocks. We use the kinematics of the \ha\ emission from HII regions and an automated modeling method to put the first-ever constraints on the encounter parameters of one of the observed systems. Besides, we roughly estimate the fraction of shocked \ha\ emission , $\text{f}_\text{shocked}$, without taking extinction into account and examine the spatial distribution of shocks. We find that close galaxy pairs have, on average, a higher shock fraction than wide pairs, and coalesced mergers have the highest average $\text{f}_\text{shocked}$. In addition, galaxy pairs with more equal mass ratio tend to have a higher $\text{f}_\text{shocked}$. Combining the dynamical models from the literature and this work, we inspect trends between $\text{f}_\text{shocked}$ and dynamical encounter parameters. Our findings are generally consistent with shocks being produced either by the direct collision of the ISM or by the chain of events provoked by the tidal impulse during the first passage.  
\end{abstract}




\section{Introduction}
\label{sec:obssys:introduction}

Galaxies have evolved in size, shape, kinematic properties, and star formation rate (SFR) over the past few billion years \citep{Bell:2004foa,Faber:2007btb,Brown:2007ca}, and galaxy mergers may have played a significant role in this evolution \citep{Toomre:1972jia,Narayanan:2008hc,Lotz:2011bua,RodriguezGomez:2016im}. In most galaxy mergers, the gravitational tidal force between interacting galaxies and the orbital energy that is converted into internal energy of the remnant can redistribute stars, gas, and dark matter particles in the phase space, leading to the transformation of the shape and kinematic properties of galaxies over time (\citealt{Toomre1977MergersConsequences}; \citealt{Schweizer:1982ex}). Moreover, interaction affects the star formation rate (SFR) both in the core and in the outskirts of gas rich galaxies \citep{Barnes:1996bna,Sanders:1996cd,Mihos:1996boa,deGrijs:2003jr,Cox:2008jj}. In case of certain encounter geometries, gas clouds of two or more interacting galaxies directly collide, and due to ram pressure and tidal stripping and shock-heating, the fraction, temperature, and density of the gas in the remnant are affected \citep{Cox:2004hp,Soto:2012hm,Vollmer:2012gr}. Studying properties of gas in interacting galaxies allows us to evaluate the comparative significance of each process in the overall evolution of galaxies. 

A case study can be the evolution of the sources of nebular emission as a merger proceeds, particularly the evolution of the fraction of shock-heated gas. Merger induced shocks in nearby tidally interacting galaxies are not only observed, but also predicted by hydrodynamical simulations of binary galaxy mergers (e.g. see \citealt{MonrealIbero:2010ko,Rich:2011is,Belfiore:2016dv} for observations and \citealt{Cox:2004hp} for simulations). As we get closer to coalescence, we expect significant shock heating to take place as a result of the mixing and collision of random gas orbits as well as the return of gas removed previously through tidal striping \citep{Cox2004GeneratingMergers}. Though, shocks have also been observed in early stage mergers \citep{Soto:2012hm}. In the earlier stages, shocks may be produced through two different modes, which we call ``tidal mode" and ``direct collision mode". In the tidal mode, each of the interacting discs experiences a strong tidal force during the first passage. The maximum of this tidal force on each galaxy depends on their separation at pericenter, as well as the mass of the companion(s). Tidal force results in a gas flow toward the center of each disc (\citealt{Mihos1994TriggeringMergers,Barnes:1996bna}; \citealt{Barton2000TidallyGalaxies}). The inflowing gas may collide with the gas at lower radii at a high enough velocity to induce shocks. The inflowing gas may also reach the core and trigger (or enhance) starburst or Active Galactic Nucleus (AGN) generating strong outflows \citep{Ellison:2011fp,Scudder2012GalaxyKpc}. These outflows produce shocks as they blow into the interstellar medium (ISM). In the tidal mode, more shocks are expected if the tidal force is stronger, so we anticipate to find higher shock emission when either the companion is more massive or the pericentric separation is smaller. On the other hand, in the direct collision mode, collision of galaxies with certain geometries can result in a fast direct encounter of gaseous clouds of the ISM in the two discs, creating widespread shocks \citep{Vollmer:2012gr,Medling:2015eh}. For both of these modes, The chain of events would begin after the first passage, but the timeline would depend on the mode. In the tidal mode, a merger-driven central starburst or AGN would produce shocks a few hundreds of Myrs after the first passage \citep{Ellison:2011fp,Scudder2012GalaxyKpc}. In the direct collision mode (e.g. in UGC 12914/5), shock production would be immediate. Observing a sample of interacting galaxies and inspecting the relationship between encounter parameters and shock fraction would put new observational constraints on these speculations.

For example, \cite{Rich:2015kf} used galaxy-wide \ha\ emission in a sample of 17 (U)LIRGS to show the relation of shock fraction and merger stage. They binned galaxies (mergers) into four groups based on their separation: isolated galaxies, wide pairs with separation between 10-100 kpc, close pairs with separation$<$10 kpc, and coalesced mergers. Removing unambiguous AGN hosts from their sample, they found that going along the sequence from isolated galaxies to coalesced mergers, an increasing fraction of \ha\ is emitted from shock-heated sources, which must be a result of increasing turbulence as merger proceeds. However, the projected separation of interacting galaxies does not necessarily represent the merger stage. An interacting pair of galaxies just after the first passage are as close to each other as a pair near coalescence. Also, two galaxies at large separation appear to be a close pair from certain viewing angles. In order to properly constrain the merger stage, and other possible encounter parameters affecting shocks, we need a dynamical model of the merger system.

Dynamical modeling constrains merger stage and other encounter parameters (e.g. pericentric distance, eccentricity, etc.) utilizing simulations that best reproduce tidal tails and bridges.  Simulations that are used for this purpose are usually collisionless N-body simulations, because incorporating gas physics extensively increases the computational demand \citep{Barnes:2009fh,Holincheck2016GalaxyGalaxies}. We often need to measure and model the line of sight velocity of tidal features, in addition to their apparent shape, to ensure the uniqueness of the model \citep{Barnes:2011kb}. As a kinematic tracer, \ha\ emission is relatively easy to measure for star-forming galaxies even in low surface brightness regions. Galaxy mergers often induce star formation both in the center and in the tidal tails  \citep{Jog:1992ct,Hattori:2004ic,Whitmore:1995fh,deGrijs:2003jr,Scudder2012GalaxyKpc}. If we can separate \ha\ emission originating at star-forming regions from shocked gas, \ha\ observations may be used not only to measure the velocity of baryons in tidal features but also to detect shocks and their distribution throughout a merger.  

\cite{Mortazavi2018HMice} proposed a method for separating shocks from star-forming regions, which only uses the kinematics and ratio of \nii\ and \ha\ emission lines, for their data did not include [OIII] and H$\beta$ lines. By separating star-forming regions from shocks, they found an excellent match between the velocity of star-forming \ha\ and cold HI gas in the Mice galaxy merger. They used the automated dynamical modeling method developed in \cite{Mortazavi:2016hv} to model the Mice using both \ha\ and HI kinematics, finding consistent results. Also, they showed that the properties of shocked \ha\ emission obtained from their method is consistent with the data from CALIFA galaxy survey, which includes [OIII] and H$\beta$ lines \citep{Wild:2014do,SanchezMenguiano:2016uw}.

Galaxy-wide nebular emission in a sample of tidally interacting galaxies can be used to improve our understanding of the physical processes during the merger. In order to find the encounter parameters of galaxy mergers, the easiest kinematic tracer is the \ha\ emission line. Additionally, the \ha\ and 
\nii\  emission lines contain valuable information about
the source of ionization, especially shocks. We can 
measure what fraction of \ha\ flux is emitted 
from shock-ionized gas, and where the shocks are spatially located. We may use the reconstructed encounter parameters and the measured shocks to investigate how shocks evolve with different merger parameters. Statistical approach would help us to isolate the effect of each parameter on shocks production mechanisms.

In this work, we present \ha\ observations of a modest sample of 22 galaxy mergers using the SparsePak Integral Field Unit (IFU; \citealt{Bershady:2004gp}) on the WYIN telescope at Kitt Peak National Observatory (KPNO). Observations of 21 systems are new to this work, while the data for the Mice galaxy merger was presented before in \cite{Mortazavi2018HMice}. We present the analysis of emission lines and shock detection for all of the observed systems and a new dynamical model for one of them. We examine correlation between encounter parameters and shocks. In \S \ref{sec:obssys:obs} we describe the observational setup, the instrument, and the target selection.  In \S \ref{sec:obssys:analysis} we present the analysis of emission lines including fitting multiple kinematic components and an MCMC code to measure the error of the fit parameters. In \S \ref{sec:obssys:ionization} we discuss how to separate emission originating at shocked gas from that arising out of star-forming regions, and we inspect the relationship between \ha\ shock fraction and encounter parameters. In \S \ref{sec:obssys:model} we present our attempts to model the dynamics of equal mass mergers in our sample, and in \S \ref{sec:obssys:fshockvsparams} we inspect the correlations between shock fraction and some of the encounter parameters. In \S \ref{sec:obssys:discussion} we discuss our results and implications about how merger encounter parameters affect merger-induced shocks. We present some notes on a couple of observed systems in the Appendix. Throughout the paper, we assume a flat cosmology with $\Omega_{m,0}=1-\Omega_{\lambda}=0.3$ and $H_0 = 70 \textrm{km } \textrm{s}^{-1} \textrm{Mpc}^{-1}$.
 
\section{Observations}
\label{sec:obssys:obs}

We observed 22 galaxy merger systems using the SparsePak IFU on the WIYN telescope at KPNO. The observations took place from March 2008 to May 2013 in five observing runs consisting of a total of 14 nights, with the dome being 
closed three full nights and parts of some other nights due to weather conditions. Our primary goal was to measure the kinematics of \ha\ and \nii\ emission lines, so we kept observing in non-photometric conditions, and occasionally thin clouds were covering our targets during the observation. Resultingly, we did not attempt to flux calibrate our data.

\subsection{The Instrument}
\label{sec:obssys:obs:inst}

The WIYN SparsePak IFU consists of a grid of 82 sparsely packed fibers each 4.687'' in diameter, covering a Field of View (FoV) of 72''$\times$71.3''. Seven sky fibers are placed at $\approx 25''$ distance from the science grid on the north and east sides \citep{Bershady:2004gp}. 

The bench spectrograph and 860 lines/mm grating blazed at 30.9$^{\circ}$ in order 2 was used, obtaining a dispersion of 0.69 \AA/pixel (FWHM) in the wavelength range of 6050-7000\AA. (R$ \sim $4500, and velocity resolution $\sim $30 km/s near the \ha\ line) Our spectral coverage is less than current and ongoing galaxy surveys such as CALIFA \citep{Sanchez:2012ku}, MaNGA \citep{Bundy:2015ft} and SAMI \citep{Croom:2012fo}. Our spectral resolution, however, is higher than CALIFA and MaNGA. In red band, CALIFA, and 
MaNGA surveys have spectral resolution, R, of 850 and 2000, respectively. 
Higher spectral resolution enables us to resolve multiple emission 
line components, usually appearing in the central regions of galaxies 
where multiple gaseous components overlap.

SparsePak is especially suitable for the purpose of finding velocity maps for 
dynamical modeling of galaxy mergers, as it has a sparsely packed grid that covers a relatively large FoV. Modeling the dynamics of interacting galaxies does not require a uniform velocity coverage on the system. Rather, velocity information that reveals the direction of rotation near the center and the large scale velocity gradient across the tidal tails and bridges is enough for dynamical modeling. Previous dynamical models of interacting galaxy pairs have used HI velocity maps with lower spatial resolution than is available with SparsePak \citep{Hibbard:1995iz,Privon:2013fs}. In this work, in order to observe each part of the targets, we mostly pointed SparsePak only once with no dithering.\footnote{In some of the coalesced systems we dithered in order to find a better coverage and better handle on the spatial distribution of outflows.} The slightly denser grid in the center of the SparsePak footprint helps to better constrain the rotation near the cores of the galaxies. 

SparsePak, however, is not ideal for inspecting shocks in galaxies. For studying shocks, full spatial coverage is more desirable as we may miss regions with extensive shocks between the sparsely placed fibers. In addition, by pointing the denser center of SparsePak toward the center of galaxies, in most systems, we introduce a bias in the galaxy-wide shock fraction toward the value near the center.

\subsection{Target Selection}
\label{sec:obssys:obs:tgtsel}

Primarily, we selected binary systems of interacting galaxies with strong tidal features. In addition, our sample includes three coalesced systems with strong tidal features, suggesting that they have also experienced a recent major merger. Tidal features are one of the best indicators of a recent merger event of rotation-supported disc galaxies. These features would not have survived if the interacting galaxies were dispersion-supported. Besides, these features can be reproduced by test particle simulations (\citealt{Toomre:1972jia}; \citealt{Wallin1992MassGalaxies}; \citealt{Dubinski:1999jg}), and their shape and velocity is sensitive to the encounter parameters, so they are utilized for reconstructing encounter parameters \citep{Barnes:2009fh,Barnes:2011kb,Mortazavi:2016hv,Holincheck2016GalaxyGalaxies}. 

All but one of the observed systems were in the footprint of Sloan Digital Sky Survey (SDSS \citealt{York:2000gn})\footnote{Arp 273, also known as the Rose Galaxies, is not in SDSS footprint. However, the Hubble space telescope (HST) image taken in 2010 resolves individual HII regions in both galaxies.}. We required both galaxies in each system to have SDSS spectroscopic data near the center showing \ha\ emission. We also required the rest of the galaxies to have blue color, so they were likely to host star-forming regions. This introduces another selection bias in our sample toward star-forming galaxies.

\begin{figure*}
\begin{center}
\includegraphics[width=1.0\textwidth]{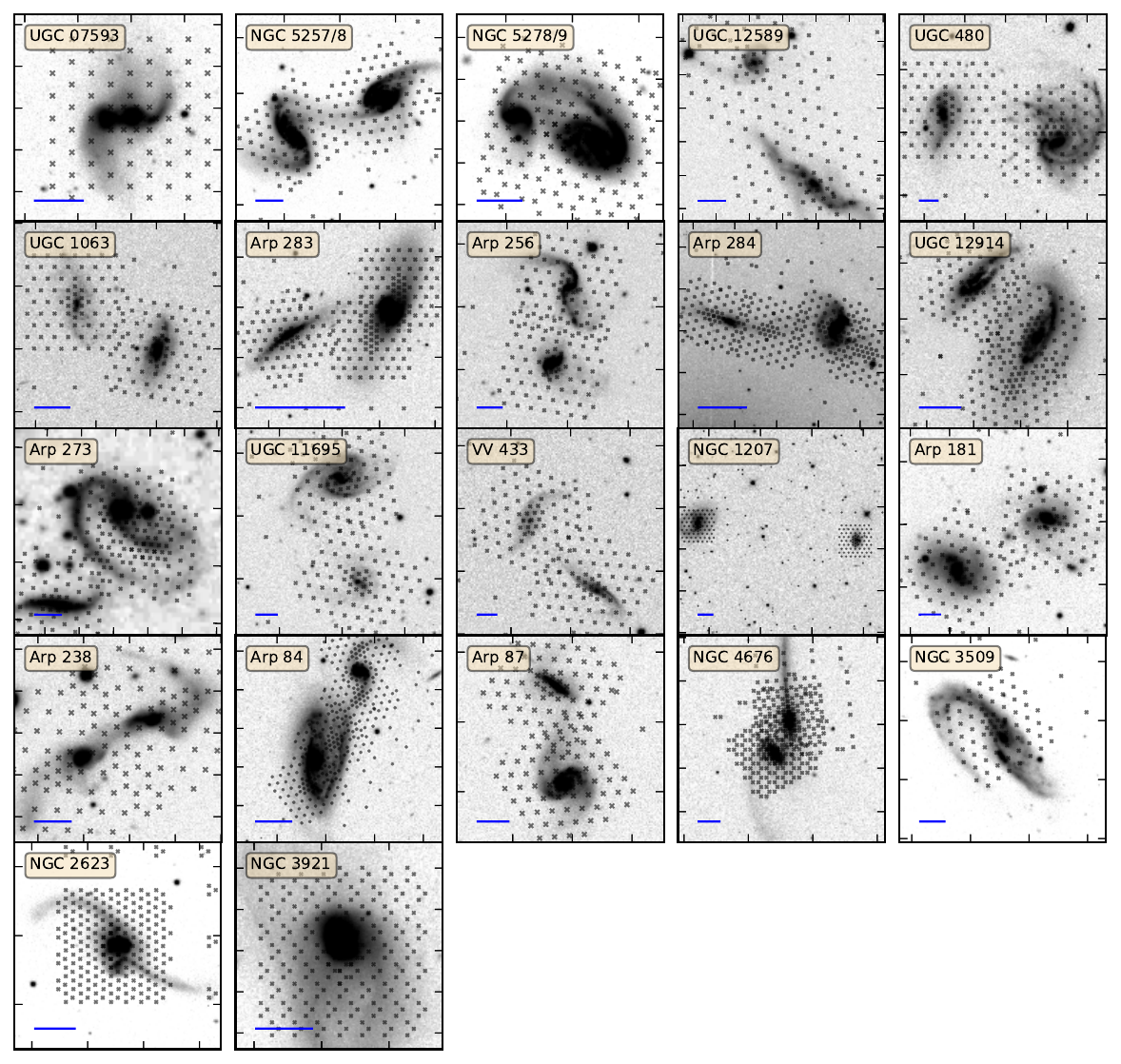}
\caption[A collage of systems observed and analyzed with SparsePak]{
A collage of systems observed with SparsePak. SparsePak pointings
used are displayed on each image with x marks showing the fiber positions. Images are scaled to fit the tiles. the blue lines show a length of 10 kpc in each panel. The images are obtained from r-band SDSS, except for Arp 273 which is from the Digitized Sky Survey (DSS).}
\label{fig:obssys:collage}
\end{center}
\end{figure*}

We used three different catalogs to find the merger systems that match our requirements. Our targets are chosen from the catalog of isolated pairs of galaxies in the northern hemisphere \citep{1985BICDS..29...87K}, the Arp atlas of peculiar galaxies \citep{1966ApJS...14....1A}, and the \textit{Galaxy Zoo} morphological classification catalog \citep{Lintott:2010gb}. In the \textit{Galaxy Zoo} catalog, we visually inspected galaxies with merger probability $>$0.5 and selected proper candidates.

Targets are selected to have an angular size that matches the Field of View (FoV) of SparsePak science grid (72''$\times$71.3''). Most of the observed systems and their tidal tails fit in two SparsePak pointings, one for each galaxy. Some smaller systems fit in one pointing, and some larger ones require three or four pointings. All of the systems analyzed in this work are shown in Figure \ref{fig:obssys:collage} along with the SparsePak grid pointings. Table \ref{tab:obssys:date} shows the sky position of the targets, observation date, redshift, number of SparsePak pointings, and exposure time on each pointing.

\begin{table*}
\centering
\begin{tabular}{l|llllll}
system  & RA 	& DEC 	& observation 	& redshift 	& \# of  			& exposure  		\\ 
name	& (deg)&(deg)				&	date				&					&	pointings	&	time (mins)	\\
\hline
UGC 12914 & 0.4171 & 23.4898 & Oct 2012 & 0.0146 & 3 & 65, 30, 125 \\
Arp 256 & 4.7104 & -10.3693 & Oct 2012 & 0.0272 & 2 & 40, 65 \\
VV 433 & 9.8322 & 13.1064 & Oct 2012 & 0.0353 & 2 & 60, 105 \\
UGC 480 & 11.6472 & 36.3286 & Oct 2012 & 0.0374 & 2 & 97, 90 \\
UGC 1063 & 22.2881 & 11.1360 & Oct 2012 & 0.0193 & 2 & 65, 65 \\
Arp 273 & 35.3778 & 39.3660 & Oct 2012 & 0.0251 & 3 & 37, 50, 32 \\
NGC 1207 & 47.0034 & 38.3769 & Oct 2012 & 0.0160 & 2 & 50, 25 \\
NGC 2623 & 129.6001 & 25.7545 & Mar 2008 & 0.0185 & 3 & 185, 86, 90 \\
Arp 283 & 139.3624 & 41.9970 & Oct 2012 & 0.0060 & 3 & 55, 65, 95 \\
Arp 181 & 157.1193 & 79.8182 & May 2013 & 0.0326 & 1 & 30, 82 \\
NGC 3509 & 166.0981 & 4.8286 & Mar 2008 & 0.0257 & 1 & 110 \\
Arp 87 & 175.1850 & 22.4379 & May 2013 & 0.0237 & 2 & 25, 58 \\
NGC 3921 & 177.7786 & 55.0788 & Mar 2008 & 0.0197 & 3 & 120, 120, 120 \\
UGC 07593 & 187.0612 & 44.4532 & Apr 2012 & 0.0230 & 1 & 95 \\
NGC 4676 & 191.5443 & 30.7271 & Mar 2008 & 0.0220 & 4 & 150, 120, 120, 120 \\
Arp 238 & 198.8870 & 62.1269 & May 2013 & 0.0308 & 2 & 25, 35 \\
NGC 5257/8 & 204.9805 & 0.8354 & Apr 2012 & 0.0227 & 2 & 85, 90 \\
NGC 5278/9 & 205.4237 & 55.6722 & Apr 2012 & 0.0252 & 2 & 65, 65 \\
Arp 84 & 209.6492 & 37.4391 & May 2013 & 0.0116 & 4 & 105, 30, 35, 30 \\
UGC 11695 & 318.0418 & -1.4857 & Oct 2012 & 0.0323 & 2 & 65, 125 \\
UGC 12589 & 351.2615 & 0.0096 & Oct 2012 & 0.0338 & 2 & 45, 125 \\
Arp 284 & 354.0750 & 2.1557 & Oct 2012 & 0.0093 & 4 & 39, 35, 55, 121 \\
\end{tabular}
\caption[list of observed systems and their characteristics]
{List of observed systems sorted by the date of observation. Redshifts are measured in this work. The number of pointings indicates how many times we had to move the SparsePak IFU to cover both galaxies and their tidal tails. Exposure time is provided for each pointing in the time order. At first, we planned to reach continuum S/N$\approx$5 based on SDSS r-band image, but as part of the observing nights were lost due to weather conditions, in favor of observing more systems, we reduced the exposure times when strong emission lines were seen in the first few exposures.}
\label{tab:obssys:date}
\end{table*}

\subsection{Observation Setup}
\label{sec:obssys:obs:setup}

In order to determine the exposure time required at each pointing, we estimated the continuum flux density near the \ha\ ($\lambda 6563$ \AA) line using the SDSS r-band images. The surface brightness in the faint tidal features of our systems ranged from 8.4$\times10^{-19}$ to 3.8$\times10^{-18}$ erg s$^{-1}\text{cm}^{-2}\text{\AA}^{-1}\text{arcsec}^{-2}$. We planned to have exposure time required to achieve S/N$\approx$5 in the continuum in fibers placed on the tails. However, we lost part of our observing time due to bad weather conditions, so in order to observe more systems, we reduced the exposure time when we could see strong \ha\ lines in outskirt fibers in the first couple of exposures. Table \ref{tab:obssys:date} shows the number of pointings and the exposure time applied on each pointing. 
 
For accurate wavelength calibration, CuAr and ThAr calibration lamps were used either before or after the science exposures on most pointings. We made three or more science exposures at each pointing to correct for the night sky lines 
and the cosmic rays. We took dome and twilight flats and zero exposures at the beginning/end of the night. Sky spectra was obtained simultaneously with either the seven SparsePak sky fibers or science fibers placed on blank sky. If the outskirts of the galaxies were far so that they covered the sky fibers, we oriented the IFU to put some of the science fibers on blank sky.

\section{Emission Line Analysis}
\label{sec:obssys:analysis}

We did not make any attempt to fit the stellar model to the continuum because the tidal tails are usually too faint. The goal of theses observations were to obtain the kinematics of 
\ha\ and \nii\ emission lines throughout the systems, including the faint tidal tails. Thus, we only fit emission line models to the combination of the three lines of \ha[$\lambda6563$] and \nii[$\lambda6583$, $\lambda6548$] , which we will call \ha-\nii\ triplet from now on. The absence of stellar model affects the \ha\ line flux measurements because we do not take the underlying \ha\ absorption into account. Our measurement of \ha\ flux is underestimated, so our \nii/\ha\ is overestimated. We will discuss the effect of this in \S \ref{sec:obssys:analysis:absorption}.

The relatively high spectral resolution of our data allows us to explore fitting more than one emission line component. This step is essential for separating the shocks using the velocity dispersion of emission lines. Sometimes, two gaseous components with different line of sight velocities lay on the same fiber, and their emission line profiles blend. One-component fit may measure a velocity that is offset from the velocity of both components. It may also measure a velocity dispersion that is broader than the velocity dispersion of individual 
blending components. In the following sections
we describe the method we used for fitting one and two velocity components and estimating the errors of the fit parameters. We also address the issue of the underlying \ha\ absorption.

\subsection{One-Component Fit}
\label{sec:obssys:analysis:one}

We first fit a triple Gaussian (three Gaussian
functions with the same free velocity and velocity dispersion, $\sigma$, and free normalization factor, \ha\ flux and \nii\ fluxes, with centers separated by the wavelength difference between \nii$\lambda$6549, \ha$\lambda$6563, and \nii$\lambda$6585) and a straight line with a free slope representing the background. We use the theoretical value of 2.95 for \nii$\lambda6583.46$/\nii$\lambda6548.05$ ratio \citep{Acker:1989vj}. Along with the slope and the offset of the background we fit a total of 6 parameters. The fit range is from 6518\AA\ to 6608\AA\ in the target rest frame. For velocity dispersion measurement, we subtract the intrinsic FWHM of our SparsePak observations (1.5 \AA) from the fit FWHM of the lines in quadrature.
 
We use a Markov Chain Monte-Carlo (MCMC) method to estimate the uncertainty of the fit parameters. The initial MCMC step is determined with the least square minimization. We use the \textsc{emcee} \textsc{python} package, varying the 6-dimensional parameter with the following prior constraints.
 
 \begin{equation}
 \begin{split}
 1.5 \leq \text{slope} \geq -1.5 \\
\text{H}\alpha\text{ flux} \geq 0 \\
\text{\nii}\lambda6583.46\text{ flux} \geq 0 
\\
 600\text{ km/s}\geq\text{velocity}\geq-600\text{ km/s} \\
14 \text{\AA} \geq \text{FWHM}\geq 1 \text{\AA} 
\label{eq:obssys:prior}
\end{split}
\end{equation}
 
The velocity here is with respect to the reported redshift for each system. The log of the likelihood function is
 \begin{equation}
 \ln{p(y_i|x_i,\sigma_i,\text{model})}=\sum \left[\frac{(\text{model}(x_i)-y_i)^2}{\sigma_i^2}+\log{\frac{1}{2\pi\sigma_i^2}}\right]\text{,}
 \label{eq:obssys:likelihood}
 \end{equation}
where $y_i$ and $x_i$ are the spectrum data points, and $\sigma_i$ is the error of the flux density measurement. After we reach the equilibrium distribution, we use the 16 and 84 percentile values of a 2000-step MCMC sample for the low and high error limits.

\subsection{Two-Component Fit}
\label{sec:obssys:analysis:two}

\begin{figure}
\centering
\subfloat[]{\label{fig:obssys:onecomp}\includegraphics[width=0.45\textwidth]{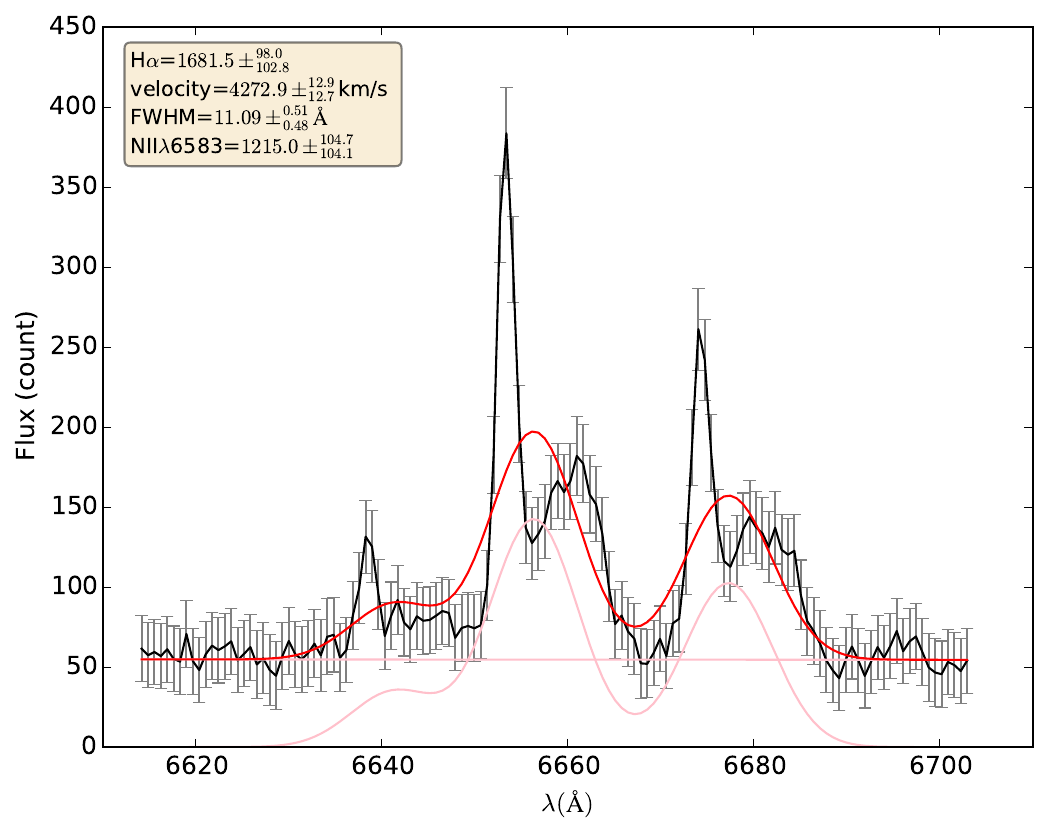}}\\
\subfloat[]{\label{fig:obssys:twocomp}\includegraphics[width=0.45\textwidth]{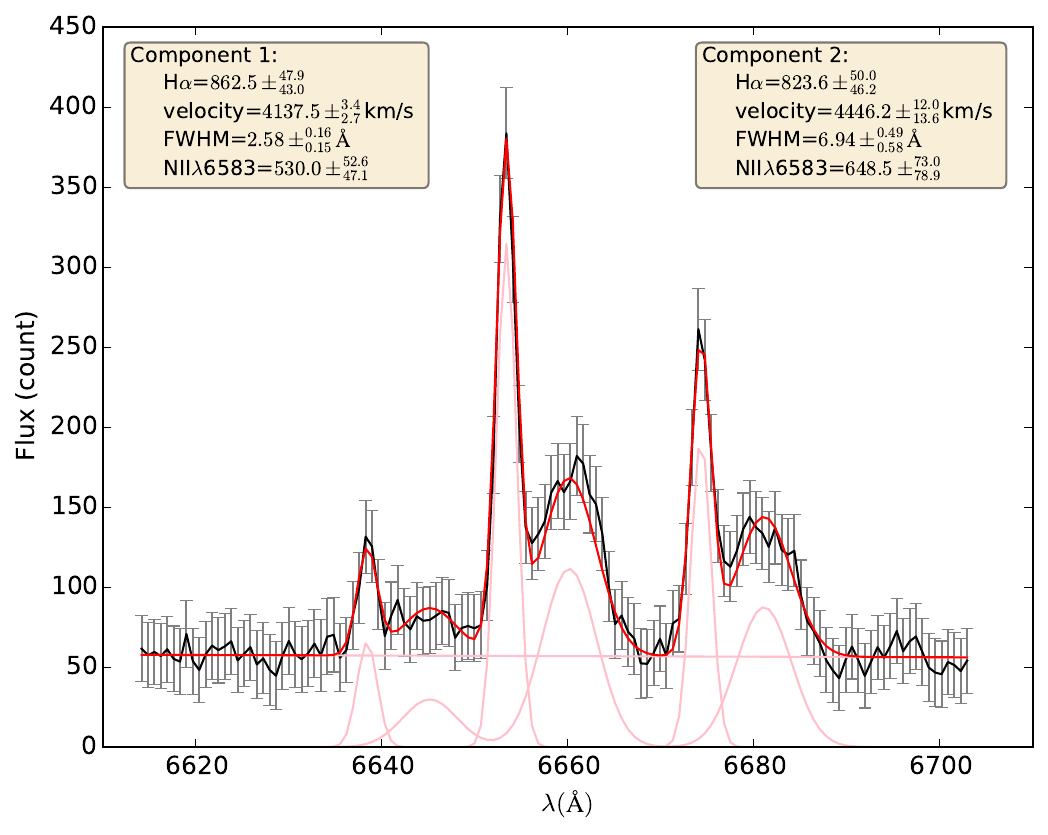}}
\caption[Comparison of single and double component fits to an emission line]{Comparison of single and double component fits to an emission line. (a) One triple-Gaussian and a background line fit
to the spectrum of a fiber in the system UGC 12914. (b) Two triple-Gaussians and a background line fit
to the same fiber. The parameters of the triple-Gaussian function and their MCMC-derived uncertainties are shown in the text box(es) in the 
upper left (and upper right) 
corner(s). It is clear that the one-component fit can not capture
all the information available in the emission line profile.}
\label{fig:obssys:one_twocomp}
\end{figure}

Visual inspection of our one-component fits makes it clear that in some of the fibers the emission lines have a more complicated profile than a simple Gaussian. Sometimes the three emission lines in the \ha-\nii\ triplet show a consistent deviation from a single Gaussian profile, indicating that the emitting gas projected into the fiber has more than one kinematic components. Figure \ref{fig:obssys:onecomp} shows an example of a fiber for which one-component triple Gaussian cannot well-describe the shape of the emission line triplet.

In order to model two-component emission lines, we fit two triple-Gaussian functions and a background line to the \ha-\nii\ triplet. This is, basically, an extra triple-Gaussian added to the one-component model. Again, we perform MCMC to estimate the error of the fit parameters. We only attempt to fit the second component when we find \ha\ flux S/N$>$3 in the one-component fit. 
The same prior constraints of Equation \ref{eq:obssys:prior} are used for both components. However, we add the following constraints to make sure that the two components have either different redshift or different FWHM. Without these constraints, two components with the same kinematic properties can share the flux arbitrary between each other, reproducing a fit that is similar to a single component fit and providing no new information about the line profile. In order to prevent MCMC walkers from wasting time in these answers, we exclude them in the prior:

 \begin{equation}
 \begin{split}
&\text{H}\alpha\text{ flux}_1 \geq \text{H}\alpha\text{ flux}_2 \\
&\text{ and} \\
&\text{velocity}_1\geq\text{velocity}_2+45\text{ km/s} \\
&\text{or }
\text{velocity}_1\leq\text{velocity}_2-45\text{ km/s} \\
&\text{or}\\
&\text{FWHM}_1\geq\text{FWHM}_2+1.5 \text{\AA} \\
&\text{or }
\text{FWHM}_1\leq\text{FWHM}_2-1.5 \text{\AA} \\
 \end{split}
\label{eq:obssys:prior2}
 \end{equation}

The likelihood function is the same as Equation \ref{eq:obssys:likelihood}. Fitting two components involves 10 parameters, so more MCMC time steps are required to achieve an equilibrium distribution. Similar to the one-component fit, after reaching equilibrium, we used the 16 and 84 percentile values of a 2000-step MCMC sample for the low and high error limits.

We use an F-test \citep{Lomax:AMzxl9ax} to determine whether a two-component model is
preferred over the one-component one. In general, increasing the number of 
parameters improves the $\chi^2$ statistics because with more parameters
the model can reproduce more subtle features in data, even if they are statistically
insignificant. The F-test utilizes both the increase in the number of free parameters and the improvement in the $\chi^2$ to determine which model is preferable. As an example, for the fit in Figure \ref{fig:obssys:one_twocomp} 
the F-test prefers the two-component fit, because the second component
significantly improves the $\chi^2$. We perform the F-test only when the \ha\ flux of the 
second component has a S/N larger than 3, otherwise the one-component fit is preferred
without performing the F-test. In our samples, about 21\% of all 956 fibers preferred a two-component fit over one-component fit.

\subsection{Underlying \ha\ absorption}
\label{sec:obssys:analysis:absorption}

By taking a straight line as the background, we are ignoring the underlying stellar \ha\ absorption. Therefore, our measurement of \ha\ flux is higher than the actual value, and our \nii/\ha\ is a higher limit. In \cite{Mortazavi2018HMice} we discussed this effect for the SparsePak spectra of the Mice galaxies (NGC 4676). For the Mice galaxies CALIFA data with a stellar continuum model is available \citep{SanchezMenguiano:2016uw}. Using CALIFA line ratios, we showed that the underlying \ha\ absorption only decrease the \logniiha\ by less than 0.2 dex (See Figure 5a of \citealt{Mortazavi2018HMice}). We showed that even with this overestimation we find shocks in the same regions of the Mice system as they were found in the CALIFA data \citep{Wild:2014do}.

As it will be discussed in \S \ref{sec:obssys:ionization:shocks}, we use \logniiha\ for shock detection, so we want to minimize the effect of the underlying \ha\ absorption on our measurement of \logniiha. In order to do so, for the purpose of our shock analysis, we remove fibers with \ha\ EW$<7$ \AA\footnote{We
keep these fibers for velocity measurements as the underlying absorption does not significantly affect the velocity of the emission lines.}. Assuming that the typical underlying \ha\ absorption EW is $\sim$ 2 \AA, for fibers with measured \ha\ emission line EW $>7$ \AA,  the \logniiha\ is only overestimated by less $<$ 0.15 dex. This is within the typical error of 
\logniiha\ in faint components. As a result, by removing fibers with \ha\ EW $<7$ \AA, we put an approximate upper limit of 0.15 dex on the overestimation of \logniiha, which is less than the maximum overestimation of \logniiha\ in the Mice system.
 
\begin{figure}
\begin{center}
\includegraphics[width=0.45\textwidth]{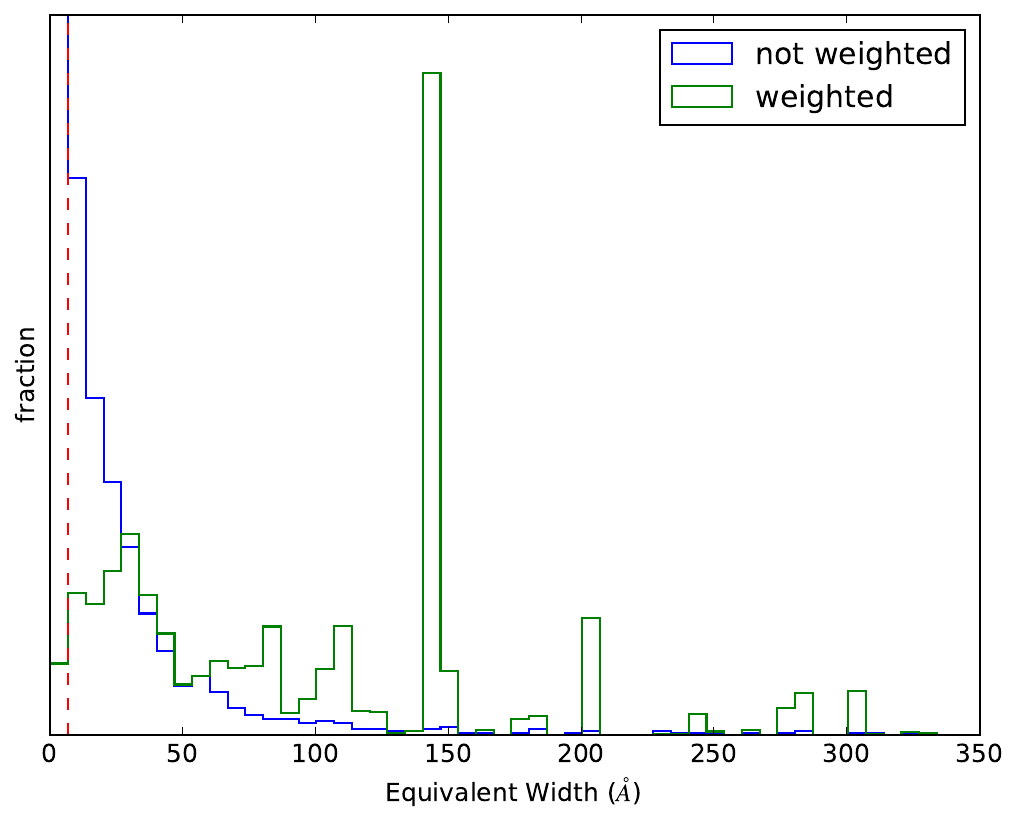}
\caption[Histogram of \ha\ EW from one-component fits]{Histogram of \ha\ EW from one-component fits. Blue and green lines show the unweighted and \ha\ flux weighted histograms, respectively. The dashed red vertical line displays the EW limit of 7\AA. This plot demonstrates that even though a large number of fibers have low \ha\ EW, most of the \ha\ flux is coming from high-\ha-EW fibers, which are not much affected by the underlying \ha\ absorption.}
\label{fig:obssys:EWhist}
\end{center}
\end{figure}

Removing the fibers with low \ha\ EW takes out a significant number of detected components with S/N $>3$ ($\approx 30\%$); however, most of the \ha\ flux in these galaxies come from the luminous fibers with high \ha\ EW, so the flux in the removed fibers are not significant, and they do not much affect the measurements of shocked \ha\ fraction. In order to demonstrate this, in Figure \ref{fig:obssys:EWhist} we show the both unweighted and \ha\ flux weighted histograms of \ha\ EW for all of the 22 observed systems. In this Figure, the cut of EW=7\AA\ is shown with the vertical red dashed line. Even though a significant fraction of unweighted histogram (blue) is to the left of the EW cut, the weighted histogram (green) clearly displays that most of the flux belongs to the fibers with EW$>$7\AA. Note that our data is not flux calibrated, so for weighting this histogram we took the total \ha\ flux in electron counts and weighted them by the exposure time. We realize that this is not a proper flux calibration method because of CCD variations and non-photometric observing conditions. We only use method for qualitative estimates similar to Figure \ref{fig:obssys:EWhist}. 

\section{Source of Ionization}
\label{sec:obssys:ionization}

\ha\ is emitted from ionized hydrogen gas, and ionization can be the result of various processes. The most common source of ionization is star formation.  The photons from these stars are typically not energetic enough to ionize metals (e.g. N and O) , and the lines usually have velocity dispersions of only a few tens of kilometers per second. The \ha\ emission from star-forming regions provide a proper kinematic tracers for dynamical modeling of disc galaxy mergers, for photo-ionization does not significantly affect the velocity of the gas. In rare situations, when the gas was already displaced from the bulk of old stars before forming new stars, the star-forming \ha\ velocity would not match the velocity of stars. We find evidence for this phenomenon in UGC 12914 (See Figure \ref{fig:appendix:ugc12914maps}). Other forms of photo-ionization such as post-Asymptotic Giant Branch (post-AGB) stars and AGNs also ionize metals along with the neutral hydrogen.  The diffuse radiation of post-AGB stars can be responsible for low-ionization emission lines like \nii$\lambda6583$, [S II]$\lambda6717,6731$, and [O I]$\lambda 6300$ \citep{Belfiore:2016dv}. The hard UV and/or X-ray radiation from AGN with energies of up to a few keVs may also produce high-ionization emission lines, such as [O III]$\lambda5003$. 
 
Ionization can also be the result of conversion of
kinetic energy into heat rather than photo-ionization. The kinetic
energy originating from the colliding gas flows produce shocks. 
The shock-heated gas cools via radiation which is itself a powerful source of ionizing photons. Photons from the post-shocked gas may travel upstream and photo-ionize the pre-shock material known as the precursor. The emission from the shock and the precursor results in high emission line ratios, particularly among the low-ionization species, such as \nii. Brightness of the radiative shocks and its precursor scales with the rate of dissipation of kinetic energy; resultingly, one expects to see higher velocity dispersion in the emission lines from brighter shocks \citep{Dopita:1995km}. Unlike photo-ionized regions around young stars, shocked gas is not a good kinematic tracer for dynamical modeling of the gravitational features like tidal tails, for the processes that produce them affect the velocity of gas at the same time. Hence, we need to separate shocks from the photo-ionized gas near young stars. In this section, we describe our method for doing so.
We investigate the how the fraction of shock-ionized gas evolve with merger stage. We also explore the correlation of shock fraction with merger mass ratio and the mass of the companion.

\subsection{Separating Shocked and Star-Forming Regions}
\label{sec:obssys:ionization:shocks}


Integral Field Spectroscopic (IFS) observations have shown that high low-ionization line ratios such as \nii/\ha\ may be found in extra-nuclear regions. Using the data from MaNGA IFU survey \citep{Bundy:2015ft}, \cite{Belfiore:2016dv} argued that most of the extra-nuclear LINER-like emission is due to ionization by post-AGB stars in old stellar populations. Post-AGB ionized gas typically has low \ha\ EW ($<$3\AA). In this work, however, we disregard all components with emission lines $<$7\AA\ in order to alleviate the issue of the underlying \ha\ absorption (see \S \ref{sec:obssys:analysis:absorption}), so it is less likely to find post-AGB ionized gas in our systems. Besides, \cite{Belfiore:2016dv} showed that high \nii/\ha\ is also indicative of shocks, and in order to distinguish them from other sources of ionization we need to look at the velocity dispersion of emission lines.

It has been shown that given high enough velocity resolution, the velocity dispersion of emission lines is also indicative of the source of ionization, particularly of shocks. \cite{MonrealIbero:2006fi} showed that in a shock-heated gas, the velocity dispersion of emission lines is correlated with
low-ionization line ratios particularly with [O I]/\ha\ and \nii/\ha.
\cite{MonrealIbero:2010ko} and \cite{Rich:2011is} confirmed these results, 
arguing that velocity dispersion can be used as an independent indicator of shocks. \cite{Rich:2011is} showed that the flux weighted histogram
of velocity dispersion of emission lines 
in their high spectral resolution 
IFU data of LIRGs reveal a bi-modality. It displays a peak
at low velocity dispersion of a few kilometers per second
corresponding to photo-ionization by stars, and a bump at
velocity dispersions higher that 100 km/s. \cite{Rich:2014ib} showed
that composite emission in the extra-nuclear regions of galaxies may originate from the combination
of shocks and star formation rather than AGN and star formation.
Based on these observations, \cite{Rich:2015kf}
proposed a limit of $\sigma<90$ km/s for velocity dispersion of 
emission from star-forming regions.
They suggested that components with 
$\sigma>90$ km/s are emitted from low velocity shocks.

\begin{figure}
\begin{center}
\includegraphics[width=0.45\textwidth]{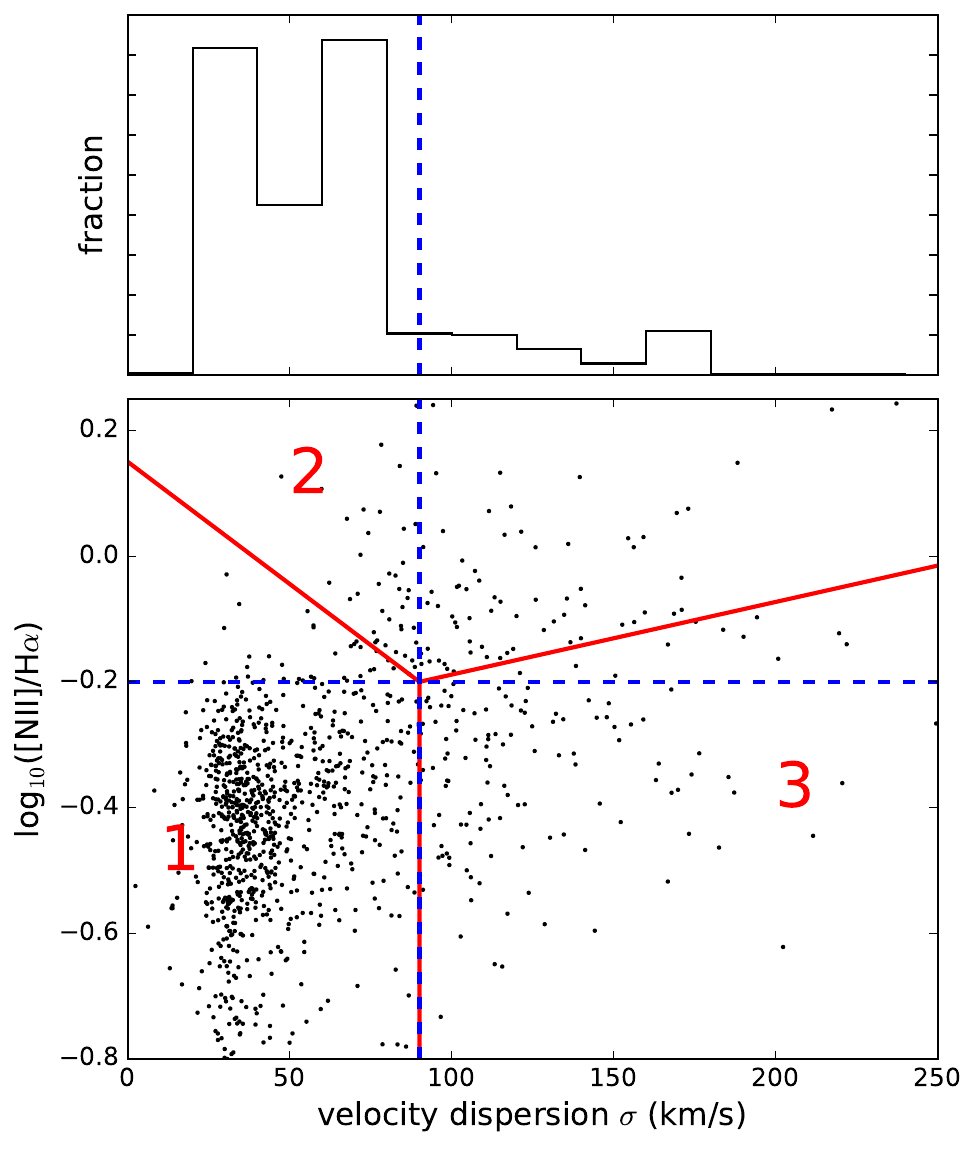}
\caption[Plot of \logniiha\ vs. velocity dispersion along with the flux-weighted histogram of velocity dispersion for all components in all systems.]
{Plot of \logniiha\ vs. velocity dispersion for all 956 components
in all galaxies in our sample. These are the components with S/N$>$3
in fibers with \ha\ EW $>7$\AA.  The concentration of points around 30 km/s is probably due to velocity resolution of our observations. Red lines are taken from \cite{Mortazavi2018HMice} and show how we determine the source of ionization. The blue dashed vertical (horizontal) line shows the limit of $\sigma=90$ km/s (\logniiha=-0.2). In this work, we take Group 1 as star-forming regions, and  Groups 2 and 3 as shocks. The panel on top shows the \ha\ flux weighted histogram of all components. Similar to some systems in \cite{Rich:2011is} and \cite{Rich:2015kf} we find a significant bumps in the \ha\ flux at high velocity dispersion ($\sim$ 160 km/s). Please note that the points in the lower panel have different \ha\ flux, while the top panel shows the fraction of \ha flux in each velocity dispersion bin.}
\label{fig:obssys:allniihasigwithhist}
\end{center}
\end{figure}

In the top plot of Figure \ref{fig:obssys:allniihasigwithhist} we show the \ha\ flux weighted histogram of velocity dispersion for all 956 high-S/N ($>$3) components in high EW fibers ($>$7\AA) in all of the observed systems. Most of the \ha\ flux from our sample is emitted at velocity dispersions below 90 km/s, which has two peaks that may correspond to normal and turbulent star formation. There is also a small bump in the \ha\ flux around the velocity dispersion of $\sim 160$ km/s.

As our observations of were carried out only in red band at a relatively high velocity resolution, similar to \cite{Mortazavi2018HMice} we combine the available information of \nii/\ha\ and width of emission line, we use plots of \logniiha\ vs. velocity dispersion to separate the shocked regions from the star-forming ones. 
In \cite{Mortazavi2018HMice} We divided the emission line components visually into three groups. The lines separating these groups are plotted in the lower panel of Figures \ref{fig:obssys:allniihasigwithhist}, as well as Figures \ref{fig:obssys:sep_sig_NHa} and \ref{fig:obssys:lratio_sig_NHa}. Group 1 has low \nii/\ha\ and low velocity dispersion. Both criteria indicate that Group 1 components are likely to be emitted from the star-forming regions. 
Group 2 components have both higher \nii/\ha\ and velocity dispersion, suggesting that shocks are their likely source of ionization. 
Group 3 components have high velocity dispersion, but relatively lower \nii/\ha. In \cite{Mortazavi2018HMice}, we took them as overlapping unresolved components of Groups 1 and 2. In this work, however, we consider Group 3 components as shocks along with Group 2. In most of the maps of \nii/\ha\ in Figure \ref{fig:obssys:niihacollage}, we find Group 3 components in extra-nuclear regions, where it is less likely that they originate from multiple unresolved velocity components. Moreover, by visually inspecting the emission lines, we confirm that most of them are indeed single high velocity dispersion components originated at shocks.

We take Group 1 as star-forming regions and Groups 2 and 3 as shocked gas. Selection of star-forming regions is fairly conservative, but for shocks one should keep in mind that high low-ionization line ratio or broad emission line can be the result of other ionizing processes such as hard AGN UV/X-ray emission and  post-AGB stars \citep{Belfiore:2016dv}. In \S \ref{sec:obssys:ionization:niihamaps} we will discuss how shocks are established as the source of ionization using the \nii/\ha\ maps of Group 2 and 3 components. 

Using the separation above we calculate the \ha\ shock fraction, $\text{f}_\text{shocked}$.  For each system, this value is obtained by dividing the sum of the \ha\ flux of Groups 2 and 3 by the total \ha\ flux of all fibers. This measurement of $\text{f}_\text{shocked}$ is an approximate estimate as the dividing lines between the groups are not rigid. Moreover, our spatial coverage is not complete, and we may have missed regions with effectively high or low shock fraction laying in between the sparsely positioned fibers. Furthermore, SparsePak has a denser grid in the center which is usually placed on the cores of the galaxies, so our measurement of \ha\ shock fraction is biased toward its value near the cores of galaxies, though most of the \ha\ flux is usually coming from the core anyway.

We should be careful not to simply interpret shock fraction as the amount of shock emission in each system. First of all, two galaxies with the same amount of shock emission and different star-formation rates will reveal different levels of shock fraction. On the other hand, we have not taken extinction into account, so in the likely case that shocked regions suffer from extinction differently compared to star-forming regions, our measure of shock fraction would not translate into the actual amount of shock emission. Though, we may assume similarity in extinction properties of shocked and star-forming regions, so the effect of extinction may be consistent, and it may not affect the general trends of shock fraction presented in this paper. A more robust measurement of shock fraction could be obtained by estimating the extinction using the available Spitzer and GALEX data in our fiber footprint similar to \cite{Smith2016AGALAXIES}. This analysis was beyond the scope of this work.

We estimate the uncertainty of $\text{f}_\text{shocked}$ ($\sigma\ {\text{f}}_{\text{shocked}}$) by bootstrap sampling of data points in the \logniiha-velocity dispersion space, based on the measured uncertainty of \logniiha\ and velocity dispersion. For each sample, we add up the \ha\ flux of Groups 2 and 3 and divide by the total \ha\ flux. The standard deviation of the \ha\ shock fraction in 100 samplings is reported as $\sigma\ {\text{f}}_{\text{shocked}}$.

Table \ref{tab:obssys:results} shows some of the quantities measured for the observed systems. 
The mass ratio of a galaxy merger can be estimated from the stellar mass ratio, assuming that the halo mass grows with stellar mass. This assumption is not accurate all the time \citep{Behroozi:2013fg}, but for the sake of simplicity we hold on to it throughout this paper. We use the Ks-band magnitude from 2MASS survey \citep{Skrutskie:2006hl} to estimate stellar mass ratio between the galaxies of each pair, so in this paper, we use the terms light ratio and mass ratio interchangeably. \footnote{We are not concerned about blending in the 2MASS photometry of our pairs. The typical spatial resolution of 2MASS survey is 4", which is smaller than the typical separation of the pairs in our sample. Besides, most of the light in our galaxies come from the core.} The projected separation between the cores of the two galaxies is measured using the redshift distance and the angular separation. In this table, we also present the minimum \ha\ EW among fibers with \ha\ S/N$>$3. We use an EW low cut of 7 \AA\ in all systems as discussed in \S \ref{sec:obssys:analysis:absorption}. In addition, we present \ha\ shock fraction ,$\text{f}_\text{shocked}$ , and its uncertainty, $\sigma\ {\text{f}}_{\text{shocked}}$ in Table \ref{tab:obssys:results}.

\begin{table*}
\begin{tabular}{l|lllllll}
system name & light ratio (Ks band) & separation (kpc) & minimum \ha\ EW(\AA) & $\text{f}_\text{shocked}$ & $\sigma\ \text{f}_\text{shocked}$ \\ \hline
NGC 1207 & 8.34 & 113.7 & 1.0 & 0.034 & 0.059 \\
UGC 480 & 2.70 & 63.9 & 1.4 & 0.213 & 0.030 \\
UGC 11695 & 8.00 & 51.5 & 4.6 & 0.239 & 0.028 \\
UGC 12589 & 3.64 & 51.4 & 2.0 & 0.239 & 0.018 \\
VV 433 & 3.12 & 50.5 & 4.1 & 0.126 & 0.058 \\
Arp 181 & 1.98 & 49.2 & 1.4 & 0.004 & 0.005 \\
Arp 273 & 2.84 & 41.7 & 0.6 & 0.304 & 0.079 \\
NGC 5257/8 & 1.17 & 37.4 & 1.5 & 0.151 & 0.023 \\
Arp 256 & 1.90 & 31.3 & 5.6 & 0.134 & 0.024 \\
Arp 87 & 1.27 & 30.8 & 1.6 & 0.412 & 0.014 \\
Arp 84 & 5.00 & 28.9 & 0.4 & 0.101 & 0.005 \\
UGC 1063 & 9.44 & 26.7 & 6.3 & 0.002 & 0.003 \\ \hline
Arp 238 & 1.51 & 22.4 & 7.0 & 0.722 & 0.073 \\
Arp 284 & 8.81 & 22.4 & 1.0 & 0.392 & 0.008 \\
UGC 12914 & 1.61 & 20.2 & 1.8 & 0.378 & 0.036 \\
NGC 5278/9 & 3.19 & 19.7 & 1.5 & 0.250 & 0.031 \\
NGC 4676 & 1.25$^*$ & 16.1 & 1.0 & 0.257 & 0.024 \\
Arp 283 & 8.43 & 11.5 & 0.7 & 0.232 & 0.059 \\
UGC 07593 & 1.81$^{**}$ & 5.6 & 0.8 & 0.401 & 0.001 \\ \hline
NGC 3509 & - & 0.0 & 3.4 & 0.012 & 0.012 \\
NGC 2623 & - & 0.0 & 0.3 & 0.895 & 0.050 \\
NGC 3921 & - & 0.0 & 1.0 & 0.702 & 0.064 \\
\end{tabular}
\caption[List of some of the measured quantities for each system.]
{List of some of the measured quantities for each system. This is sorted based on the projected separation of the centers of each pair. Horizontal lines indicate where we divide the sample into bins of wide pairs, close pairs, and coalesced systems. The projected separations are measured using the angular separation and the redshift distance.  For each system, the minimum \ha\ EW  in fibers with S/N $>3$ is presented. The highest value of the minimum EW is 7.0\AA\ (Arp 283). In order to measure a consistent shock fraction among all systems and to minimize the effect of the underlying \ha\ absorption, we remove fibers with EW $<7$\AA\ from our shock analysis. Shocked \ha\ fraction and error are shown in the last two columns. For most systems the light ratio is obtained from the 2MASS Ks-band magnitudes. $^*$ For NGC4676, no measurements were found in the 2MASS catalogue. Instead we used the stellar mass ratio from \cite{Wild:2014do}. $^{**}$ The UGC 07593 pair is not resolved in 2MASS,
so the SDSS r-band photometry is used in this table.}
\label{tab:obssys:results}
\end{table*}

\subsection{Indications for Galaxy-Wide Shocks from \nii/\ha\ Maps}
\label{sec:obssys:ionization:niihamaps}

IFS observations of systems with unambiguous AGN usually
display a gradient of emission line ratios from center to edge. Hard ionizing radiation from AGN affects the gas
in its vicinity more than the gas that is kpcs away in the disc
or above it. In the presence of an AGN in the 
center one expects to find higher \nii/\ha\ near
the center than in the outskirts
\citep{Davies:2014co,Leslie:2014dj}.
In case of shocks, on the other hand, depending on the shock producing mechanism the regions of high velocity
dispersion and  \nii/\ha\ are not necessarily near the center, but can be anywhere in the galaxy.
Resultingly, if we see no gradient of \nii/\ha\ 
toward the center, we can argue that the source of ionization is
likely to be shocks. Nevertheless, even if we find a gradient of \nii/\ha\ toward the center, it is not inevitably due to hard ionizing radiation from an AGN. If the source of shocks are the superwinds or outflows from a central processes such as starbursts or AGNs, they can also
produce a gradients in the observed \nii/\ha. The mere existence of 
a gradient in \nii/\ha\ is not an indication of an AGN.

\begin{figure*}[htbp]
\begin{center}
\includegraphics[width=1.0\textwidth]{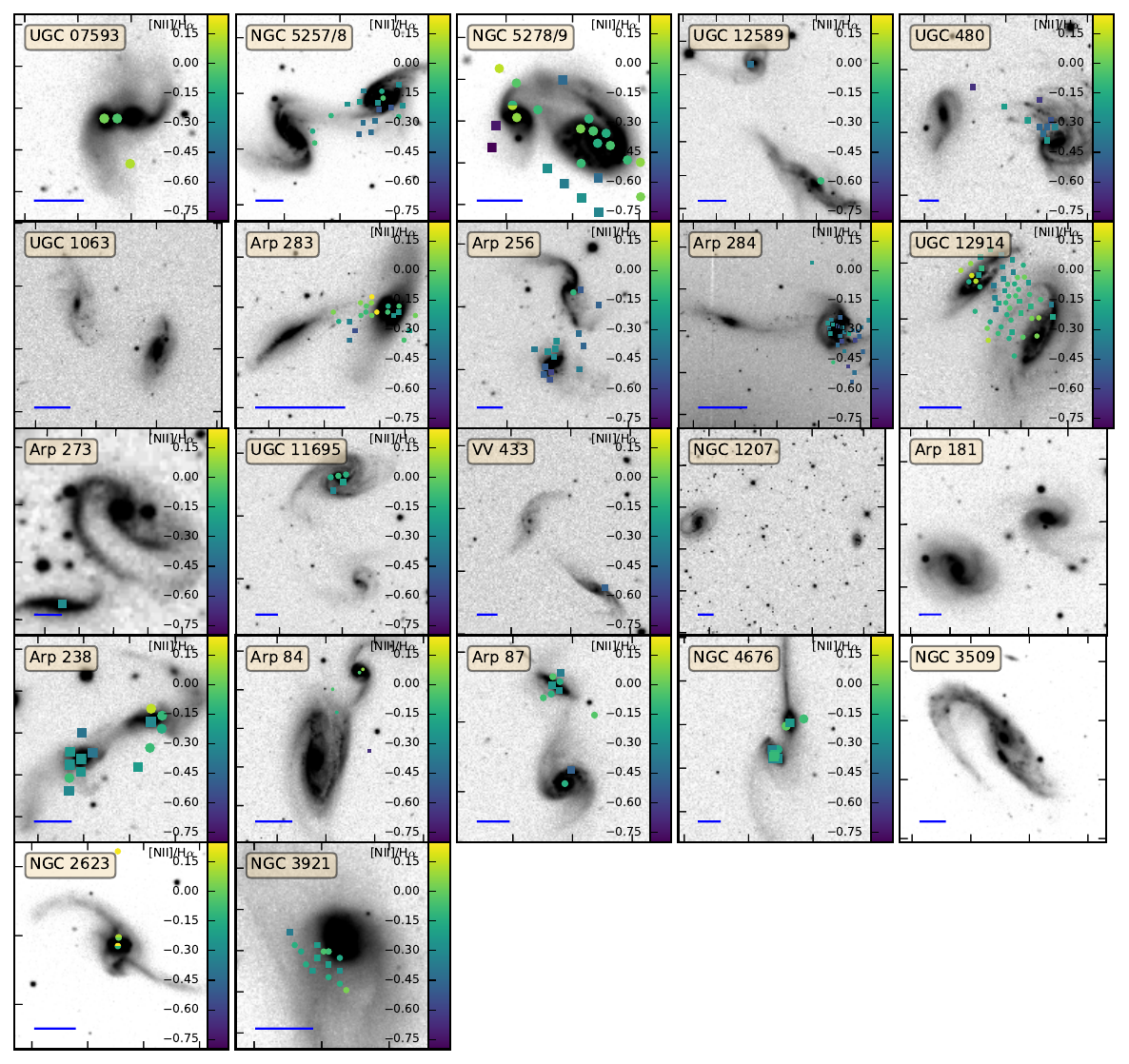}
\caption[Collage of \nii/\ha\ maps of the observed systems]{Collage of \nii/\ha
maps of Group 2 and 3 components (likely to be emitted from shock-heated gas) 
in all of the observed systems. Group 2 components are shown
with circles and group 3 components are shown with squares. The spatial distribution
of \nii/\ha\ in most of the systems suggest that these components are produced by shocks. Similar to Figure \ref{fig:obssys:collage} the blue lines show a length of 10 kpc in each panel.}
\label{fig:obssys:niihacollage}
\end{center}
\end{figure*}

Based on the arguments above, we look for indications of shocks in \nii/\ha\ maps. None of these systems are unambiguous AGNs according to the literature.\footnote{By unambiguous AGN we mean galaxies that show all AGN signatures, so there is a consensus in the literature that they are AGNs. Some galaxies like NGC 2623 are AGN candidates, but for all of them the controversy remains.} Figure \ref{fig:obssys:niihacollage} shows a collage of \nii/\ha\ maps of Group 2 and 3 components. For these maps we have implemented the \ha\ EW limit to reduce the effect of the underlying \ha\ absorption. This removed about 30\% of the components, most of them with high \nii/\ha, but they contained less that 5\% of the total \ha\ flux (Figure \ref{fig:obssys:EWhist}). In the maps of the remaining components in Figure \ref{fig:obssys:niihacollage}, we find that four systems display no Group 2 and 3 components. In 12 out of the remaining 18 systems, indications from the spatial distribution of Group 2 and 3 components and the gradient of \nii/\ha\ suggest that these components are most likely emitted from shocked gas. For example, in UGC 12914 (shocked gas fraction, $f_{\text{shocked}}$ = 37\%) these components all lie in the bridge between the galaxies, and are more likely to be the result of 
collision of gas clouds during the head-on encounter between the two discs. In Arp 284,
UGC 480, and NGC 3921 we see a cone-like structure toward the center, though
the gradient of \nii/\ha\ does not suggest a central hard ionizing source.
In the remaining six systems (UGC 07593, UGC 12589, Arp 273, VV 433, Arp 84, and NGC 2623), we only find one or two Group 2 and 3 components near the center. While AGN is a candidate for the source of ionization, it could be the result of shocks from central gas inflows and outflows, therefore the source is ambiguous in these systems.

\subsection{Shocked Gas and Merger Sequence}
\label{sec:obssys:ionization:sequence}

\begin{figure}
\centering
\subfloat[]{\label{fig:obssys:fshockseparaion_nobin}\includegraphics[width=0.45\textwidth]{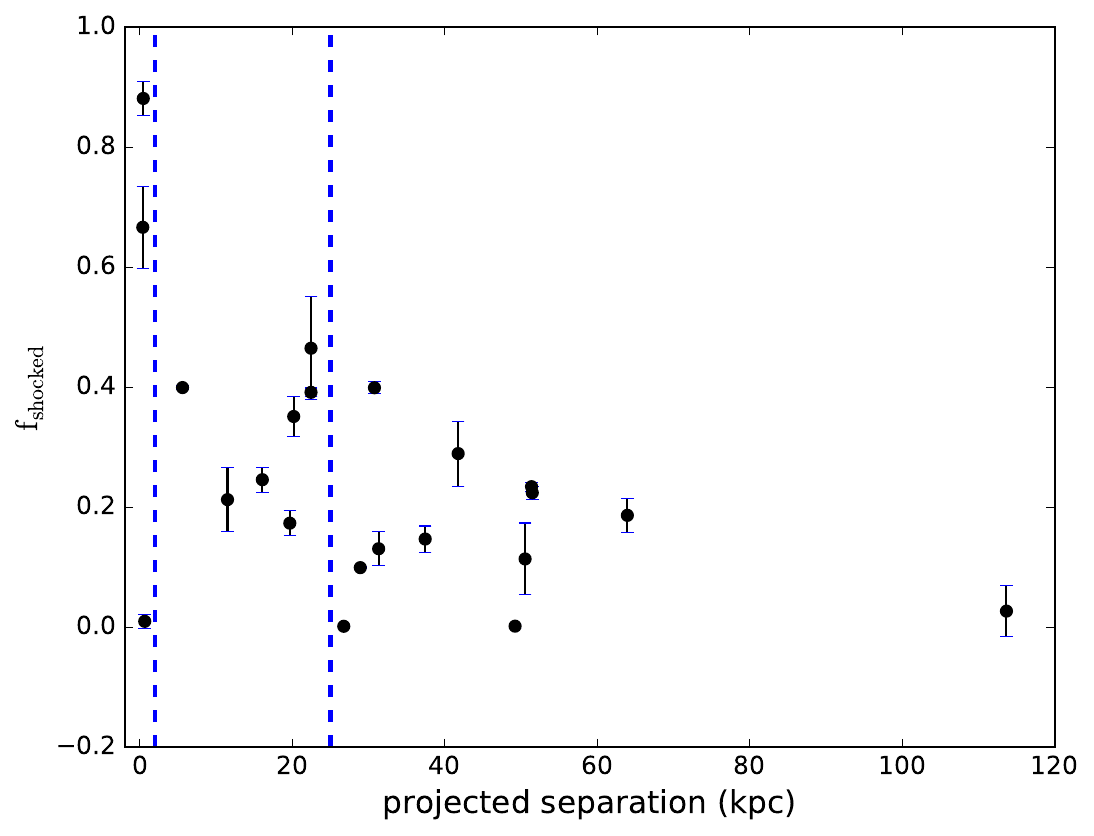}}\\
\subfloat[]{\label{fig:obssys:fshockseparaion_binned}\includegraphics[width=0.45\textwidth]{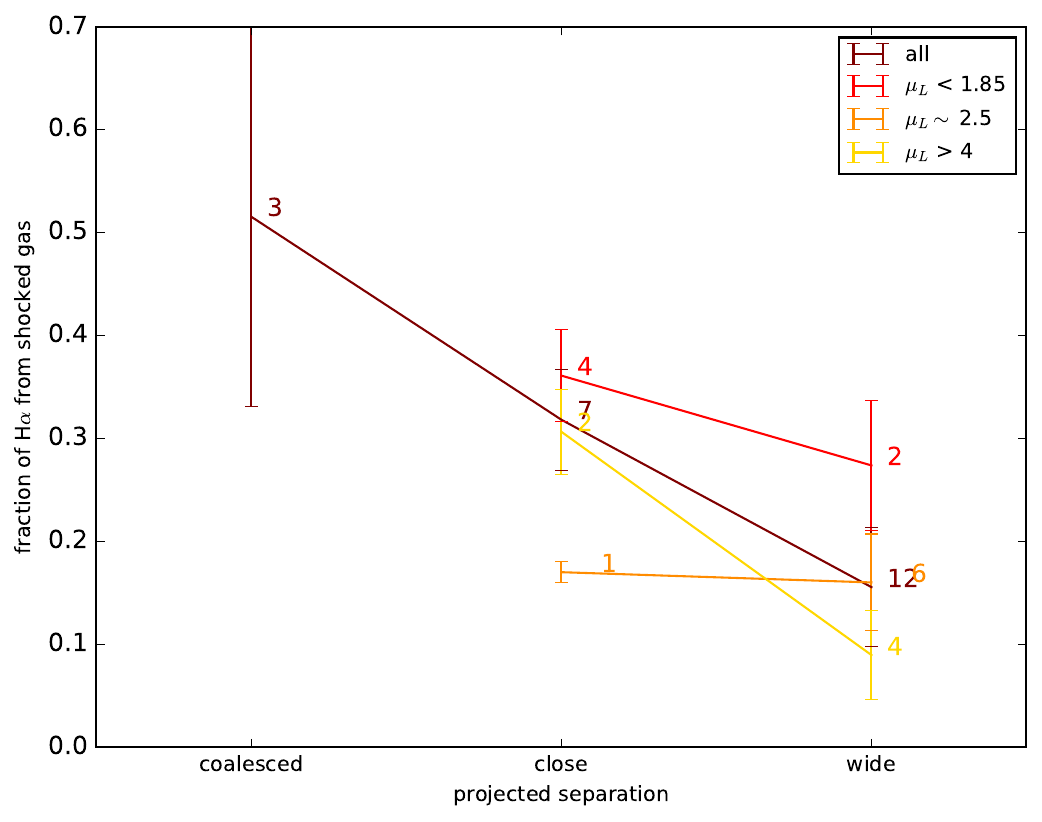}}
\caption[plot of shock fraction vs. projected separation: binned and un-binned]{(a)Plot of shock fraction vs. projected separation between the two galaxies color coded 
with difference between the velocities in the core. One can see an anti-correlation.
Close pairs display more shock fraction and wide pairs show less. two of the coalesced systems
show shock fraction higher than $>$ 80\% and one has a shock fraction of $<$ 10 \%.
Vertical dashed lines show the separation used for binning in (b).
Close pairs are the ones with separation $<$25 kpc. The rest are considered as wide pair.
The anti-correlation is clear and significant across all mass ratios. Also note that more
equal mass merger tend to have higher shock fraction. Next to each point we have written
the number of systems in that bin. }
\label{fig:obssys:fshockseparaion}
\end{figure}

Figure \ref{fig:obssys:fshockseparaion_nobin} shows the plot of \ha\ shock fraction versus projected separation between galaxies in each pair. Assuming that the separation is an indication of merger stage, close pairs are at a later stage, closer to coalescence. One may recognize the general trend of increasing shock fraction with decreasing projected separation. Similar to \cite{Rich:2015kf} we bin our systems into close pairs, wide pairs, and coalesced mergers. The close (wide) pairs are non-coalesced systems with projected separations less (more) than 25 kpc. We select the 25 kpc limit to have significant statistics in both bins. Figure \ref{fig:obssys:fshockseparaion_binned} shows the trend in these bins. In the bins of close and wide pairs, we also separate systems based on their stellar mass ratio (K-bank light ratio). Similar to projected separation, we separate galaxies into 3 bins of light ratio shown by vertical dashed lines in Figure \ref{fig:obssys:fshocklratio_nobin}: $\mu_L<1.85$ representing the equal mass mergers, $1.85<\mu_L<4$ representing major galaxy merger of non-equal mass, and $4<\mu_L$ representing minor galaxy mergers. The dividing values are chosen to have a significant number of galaxies in each bin. We can see that not only are the close pairs more shocked than the wide pairs, but also more equal mass mergers reveal a higher fraction of shocks altogether. Similar to \cite{Rich:2015kf} we find the highest shock fraction in the coalesced systems.

\begin{figure*}
\begin{center}
\subfloat[]{\label{fig:obssys:allsigmahistseparation}\includegraphics[width=0.45\textwidth]{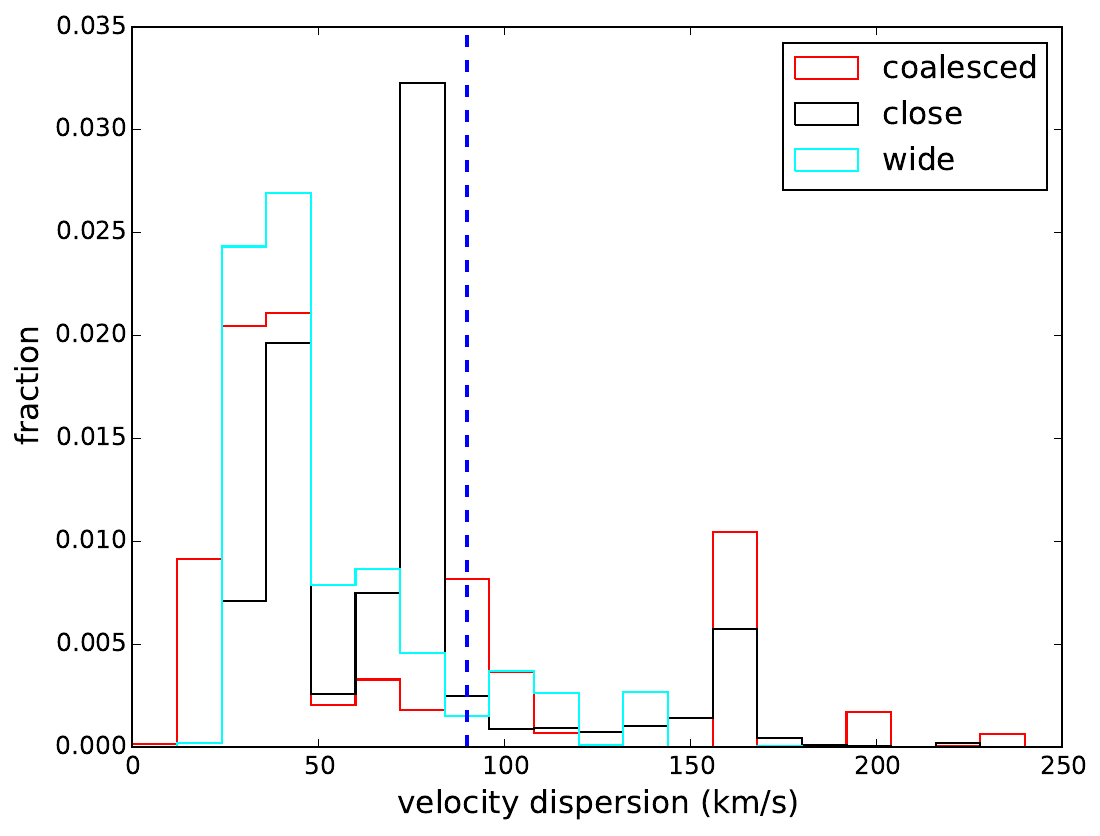}}
\subfloat[]{\label{fig:obssys:allsighistdisc}\includegraphics[width=0.45\textwidth]{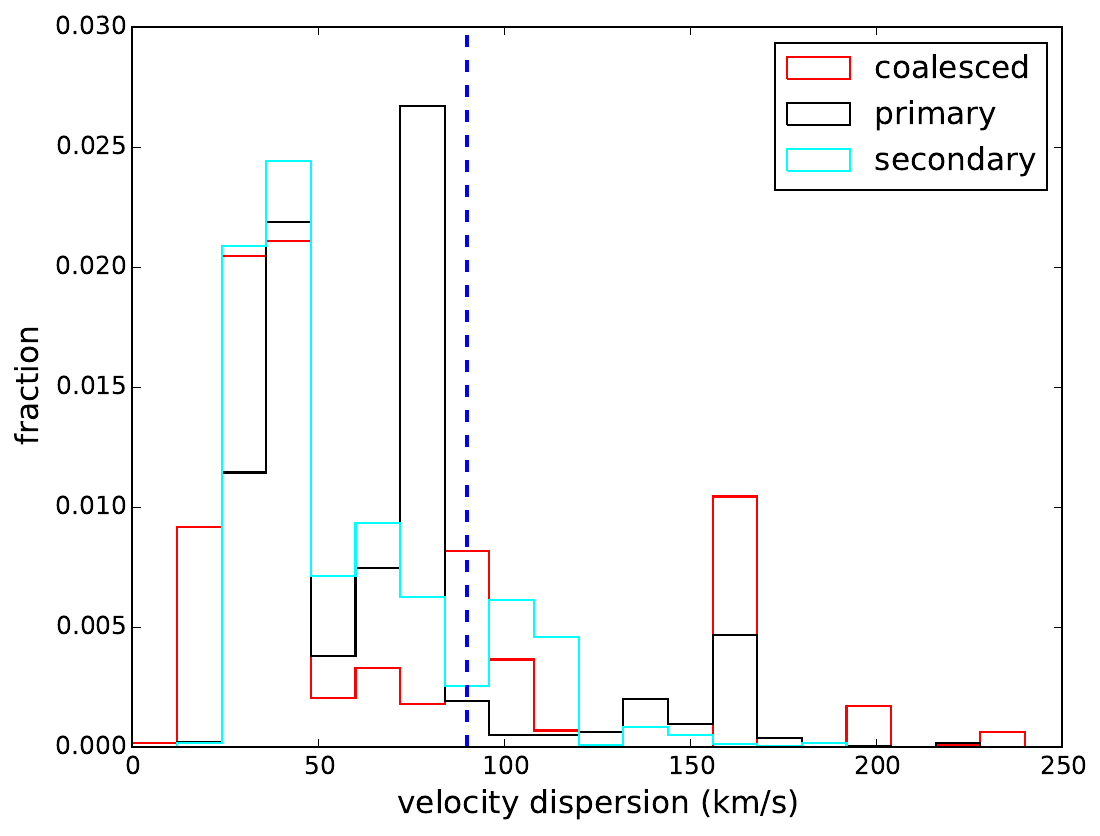}}
\caption[Histogram of velocity dispersion for all components in all systems]
{(a)\ha\ flux weighted histogram of velocity dispersion for 956 components
in all galaxies in our sample separated into coalesced and non-coalesced systems. The non-coalesced systems are also separated into close and wide pairs. Only components with \ha\ S/N$>$3
in fibers with \ha\ EW $>7$\AA are considered. The limit of 90 km/s is shown by the blue dashed
vertical line. The bi-modality is evident in  coalesced systems as well as in close pairs. Coalesced systems show a bigger bump at high velocity dispersion compared to close pairs. (b) The same histogram as (a) with the non-coalesced systems separated into the primary (more massive) and secondary (less massive) companions.  Similar to (a) the primary companions display more flux at high velocity dispersion compared to the secondary ones, but not as much as the coalesced systems.}
\label{fig:obssys:allsighist}
\end{center}
\end{figure*}

This trend is consistent with the flux-weighted histograms of velocity dispersions presented in Figure \ref{fig:obssys:allsigmahistseparation}. The coalesced systems show the highest bump at high velocity dispersion, and the close pairs emit more high velocity dispersion \ha\ flux compared to the wide pairs.

\begin{figure*}
\centering
\includegraphics[width=0.70\textwidth]{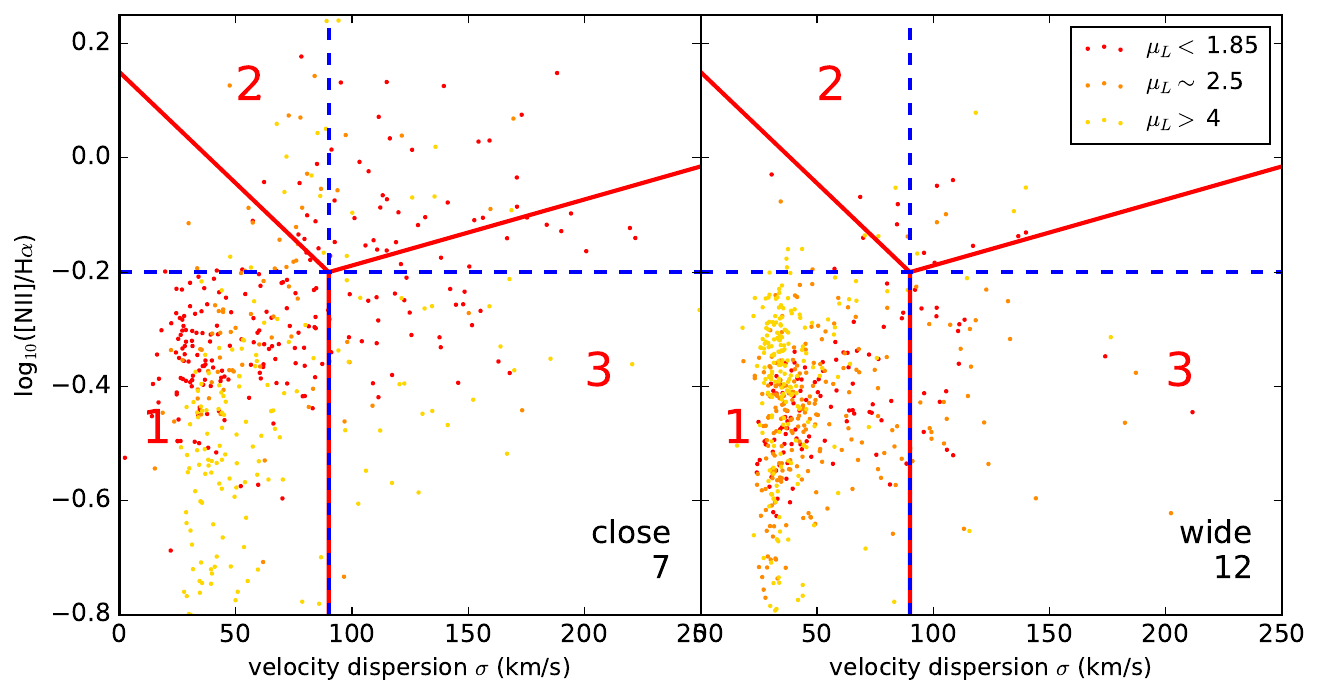}
\caption[Plot of \nii/\ha\ vs. velocity dispersion for different separations.]
{Plot of \nii/\ha\ vs. velocity dispersion for all components in wide pairs (right) close pairs (left). The number of galaxies in each bin is shown on the lower left corner of each panel. We see that in the close pairs, especially in more equal mass mergers ($\mu_L<1.85$; red points) there is a higher fraction of components in Groups 2 and 3 confirming the results of Figure \ref{fig:obssys:fshockseparaion_binned}.  Even in the Group 1 section which corresponds to star-forming regions one can see that components generally have higher \logniiha\ and velocity dispersion in the close pairs, indicating that the environment of star formation is also more turbulent. For the close and wide pairs, components of systems with different mass ratio bins are plotted with different colors. In close pairs more equal mass merger components have a significantly higher \nii/\ha. We will discuss this in \S \ref{sec:obssys:ionization:massratio}.}
\label{fig:obssys:sep_sig_NHa}
\end{figure*}

Another way to see this phenomenon is to look at the plot of \logniiha\ vs. velocity dispersion, separating components from the close and wide pairs as in Figure \ref{fig:obssys:sep_sig_NHa}. We see a significant increase in the fraction of components in the regions of Groups 2 and 3 in the close pairs compared to the wide pairs, a signature of enhanced shock ionization, which confirms the results of Figure \ref{fig:obssys:fshockseparaion_binned}. However, this plot provides extra information; we see that compared to the wide pairs in the close pairs not only is the percentage of components in Groups 2 and 3 higher, but also the components in Group 1 generally reveal a higher \logniiha\ and velocity dispersion, indicating that the environment of the ionized gas in the close pairs is more turbulent even in star-forming regions. Different light ratio bins are also separated in Figure\ref{fig:obssys:sep_sig_NHa} with red, orange, and yellow points referring to equal-mass, non-equal-mass major, and minor mergers, respectively. One can see that particularly in the close pairs more equal mass components (red points) occupy a slightly higher region compared to the other two light ratio bins. This will be discussed in \S \ref{sec:obssys:ionization:massratio}

We should keep in mind that a small projected separation does not necessarily imply a late stage merger. There are other factors than can affect the projected separation. 
For better constraints on the actual merger stage we need to model the dynamics of the merger. We will discuss this in \S \ref{sec:obssys:discussion}.

\subsection{Shocked Gas and Merger Mass Ratio}
\label{sec:obssys:ionization:massratio}

One of the benefits of having a sample of early stage mergers before coalescence is that we can explore how properties of the companion affect the other galaxy. We can find out about the properties of isolated galaxies from their cores which, supposedly, have not lost much of its stellar mass, morphology, and kinematics after the first passage. One of the properties that is relatively easy to measure for non-coalesced mergers is mass ratio. 

\begin{figure}
\centering
\subfloat[]{\label{fig:obssys:fshocklratio_nobin}\includegraphics[width=0.45\textwidth]{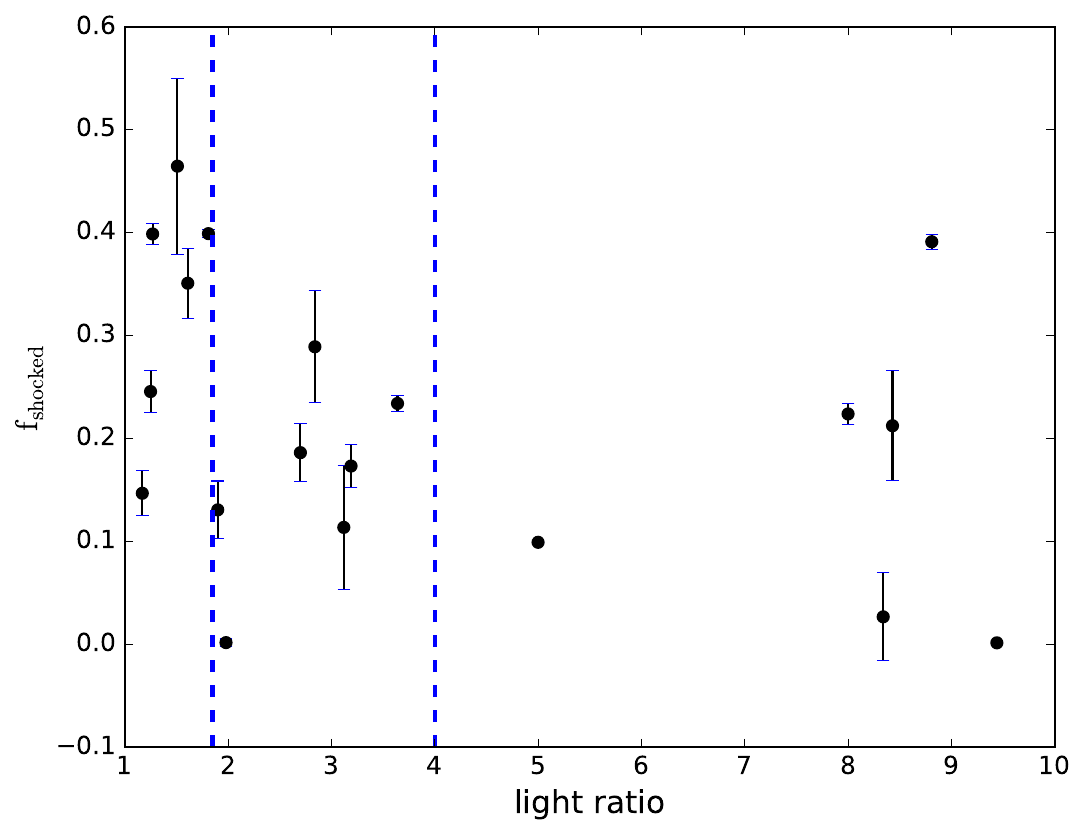}}\\
\subfloat[]{\label{fig:obssys:fshocklratio_binned}\includegraphics[width=0.45\textwidth]{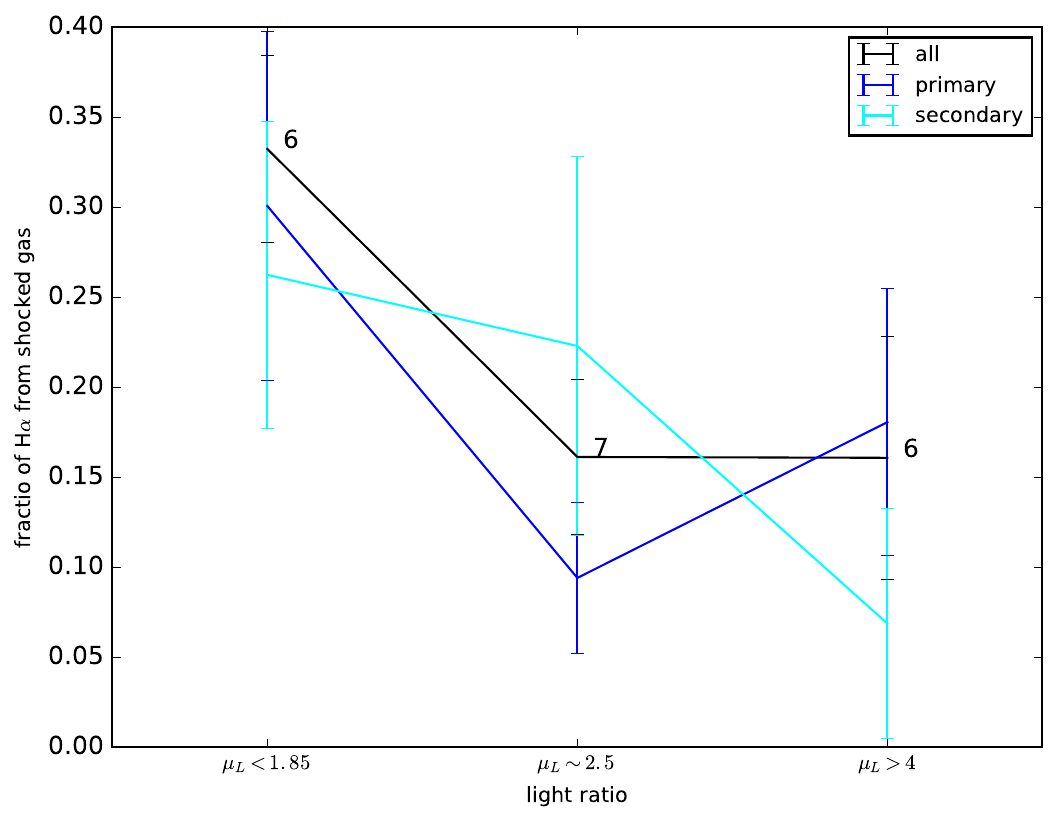}}
\caption[Plot of shock fraction vs. light ratio: binned and unbinned]
{(a) Plot of shock fraction vs. light ratio in the non-coalesced systems of our sample.
Vertical dashed lines show the light ratio bins used for binning systems in (b). (b) The trend of average shock fraction in light ratio bins. The trend for the primary, secondary, and both galaxies are shown separately. Equal mass mergers have a higher overall ionization fraction compared to the other two bins. However one should be cautious about these results because of low statistics.
The difference between shock fraction in the primary and secondary galaxies is not 
significant in these bins.}
\label{fig:obssys:fshock_lratio}
\end{figure}

Figure \ref{fig:obssys:fshocklratio_nobin} displays the plot of shock fraction vs. light ratio, $\mu_L$, in the non-coalesced pairs of our sample. Figure \ref{fig:obssys:fshocklratio_binned} shows the average shock fraction in each bin. Average shock fraction in both galaxies as well as in the primary and the secondary galaxies are shown separately. It is slightly but significantly enhanced in more equal mass mergers; though, we should be cautious about this interpretation as in such low statistics either removing one or two systems or slightly changing the limits of the bins could affect the results. Furthermore, in Figure \ref{fig:obssys:fshocklratio_binned} the error-bars are too large to provide a robust Conclusion about the difference of shock fraction in the primary (more massive) vs. secondary (less massive) companions. Though it appears that in minor mergers the shock fraction in the primary galaxies are significantly higher than the secondary ones. 

\begin{figure*}
\centering
\includegraphics[width=1.0\textwidth]{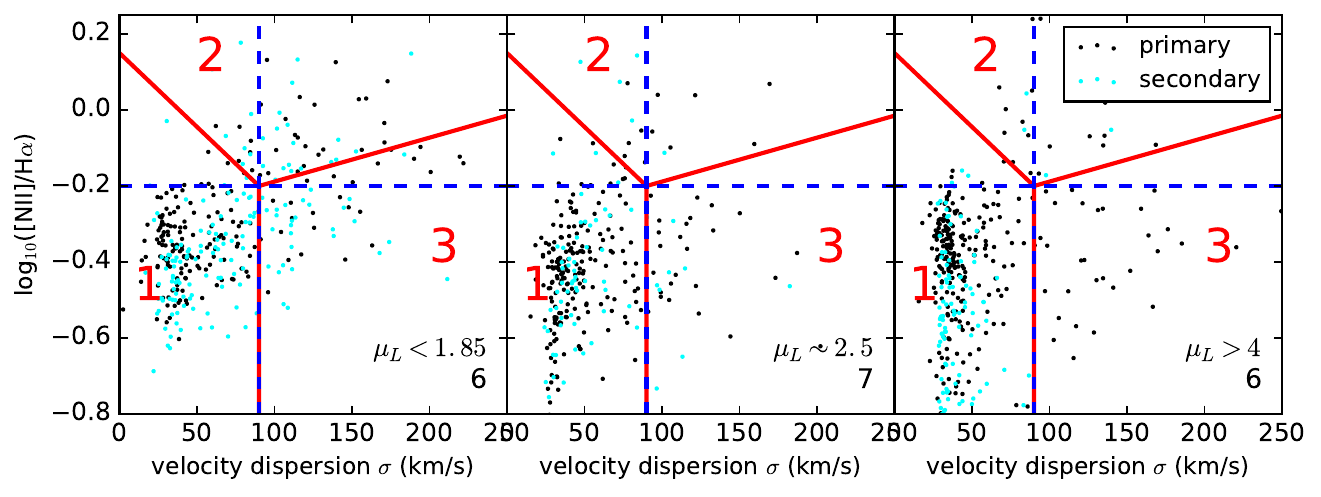}
\caption[Plots of \nii/\ha\ vs. velocity dispersion for different mass ratio bins]
{Plots \logniiha\ vs. velocity dispersion for components in different mass ratio bins: equal mass mergers
(left) $\mu_L<1.85$, non-equal major mergers (middle) $1.85<\mu_L<4.0$, and minor majors (right) $\mu_L>4.0$. The number of systems in the bins is written in lower right corner of each panel. As we we go from right to left, one can see that not only is the fraction of points in Groups 2 and 3 increasing, but also the points in Group 1 reveal a generally higher \logniiha\ and velocity dispersion. The components of the primary and secondary galaxies are plotted with black and cyan points, respectively. In the minor mergers (right panel), on average, the components of the primary galaxies have a significantly higher \logniiha\ compared to the secondary ones. }
\label{fig:obssys:lratio_sig_NHa}
\end{figure*}

Figure \ref{fig:obssys:lratio_sig_NHa} shows the components in light ratio bins on plots of \nii/\ha\ vs. velocity dispersion. Similar to Figure \ref{fig:obssys:sep_sig_NHa}, as we go from right to left not only does the fraction of points in Groups 2 and 3 increase but also the points in Group 1 reveal a slightly higher \logniiha\ and velocity dispersion. This suggests that the environment of more equal mass mergers is, on average, more turbulent even in star forming regions. Comparing the panels of Figure \ref{fig:obssys:sep_sig_NHa} we find that this effect is stronger in the close pairs compared to the wide pairs, for in the panel of close pairs, more equal mass mergers (red points) reveal a significantly higher \logniiha\ and velocity dispersion. In Figure \ref{fig:obssys:lratio_sig_NHa} black and cyan points display the components of the primary and secondary companions, respectively. In the minor merger (right panel) the black points  appear to spread significantly higher than the cyan points even in the Group 1 region, revealing a slightly higher \logniiha. Consistent with Figure \ref{fig:obssys:fshocklratio_binned}. This suggests that the environment of the primary companions are more violent that the secondary ones in minor mergers. We will discuss these results in \S \ref{sec:obssys:discussion}. 

\section{Encounter Parameters of Equal Mass Mergers - Result}
\label{sec:obssys:model}


All of the observed systems display strong elongated tidal features, which suggest that most of non-coalesced pairs are between the first and second passages. We use the method described in \cite{Mortazavi:2016hv} for measuring the encounter parameter of galaxy mergers. This method uses \textsc{identikit} \citep{Barnes:2009fh,Barnes:2011kb}, which is a software package for modeling the dynamics of interacting pairs of disc galaxies.
\cite{Mortazavi:2016hv} only tested equal mass \textsc{identikit} models, so in this work we only apply this method on the systems in the first light ratio bin in \S \ref{sec:obssys:ionization:massratio} with Ks-band light ratio $\mu_L\leq$ 1.85.

We often need to measure and model the line of sight velocity of these tidal features, in addition to their apparent shape, to ensure the uniqueness of the model \citep{Barnes:2011kb}. Collisionless stars are the ideal components to match with collisionless N-body simulations such as \textsc{identikit}, but it is often hard to measure their velocity in faint tidal tails. 
Another option is cold neutral gas, 
though detecting cold gas at high enough resolution is also relatively expensive.
Nebular emission, on the other hand, is relatively easy to measure for star-forming galaxies even in low surface brightness regions, and galaxy mergers usually enhance star formation in both the center and the tidal features \citep{Jog:1992ct,Hattori:2004ic,Whitmore:1995fh,deGrijs:2003jr}. 
However, it is important to separate star-forming regions from shocks before using the \ha\ derived velocities, for shocked gas is produced through process that significantly change their velocity compared to the bulk of the baryons \citep{Sharp:2010jl}.

\textsc{identikit} evaluates the match between model and data by calculating a parameter called ``score''. In order to calculate the score we place phase space boxes on the tidal features
of the interacting galaxies. \textsc{identikit} calculates the score for each model based on the number of test particles that populate these boxes. In this work, we use the same algorithm as in \cite{Mortazavi2018HMice} for placing the boxes, and their size are determined by the diameter of the SparsePak fibers ($\approx 5''$).


\begin{table}
\begin{tabular}{l|lll}
system name 	& light	& separation  	& previous models 	\\
			& ratio	& (kpc)		&				\\\hline
UGC 07593 & 1.81 & 5.6 & ---  \\
NGC 5257/8 & 1.17 & 33.8 &  \text{\cite{Privon:2013fs}}\\
UGC 12914 & 1.61 & 14.8 &  \text{\cite{Vollmer:2012gr}}\\
Arp 238 & 1.51 & 19.7 &  --- \\
Arp 87 & 1.27 & 2.5 & ---  \\
\end{tabular}
\caption[List of the systems we attempted to model using our method]
{List of the systems we attempted to model using our method. The light ratios are obtained from 2MASS Ks band, except for UGC 07593 which is not resolved in the 2MASS Ks survey, and we used SDSS r-band light ratio instead.}
\label{tab:obssys:model}
\end{table}

Table \ref{tab:obssys:model} shows the systems we selected for 
dynamical modeling. Here we exclude NGC 4676, the Mice, which we modeled and discussed extensively in \cite{Mortazavi2018HMice}. The systems in Table \ref{tab:obssys:model} are the most equal mass systems based on their 2MASS K-band magnitude \citep{Skrutskie:2006hl}, which is used as a crude estimate of stellar mass. We select the galaxies with a light ratio less that 1.85, and make an attempt to model them with equal mass \textsc{identikit} models. For two of these systems we found prior dynamical models in the literature, which are mentioned in Table \ref{tab:obssys:model}.



\begin{figure*}
\centering
\includegraphics[width=1.0\textwidth]{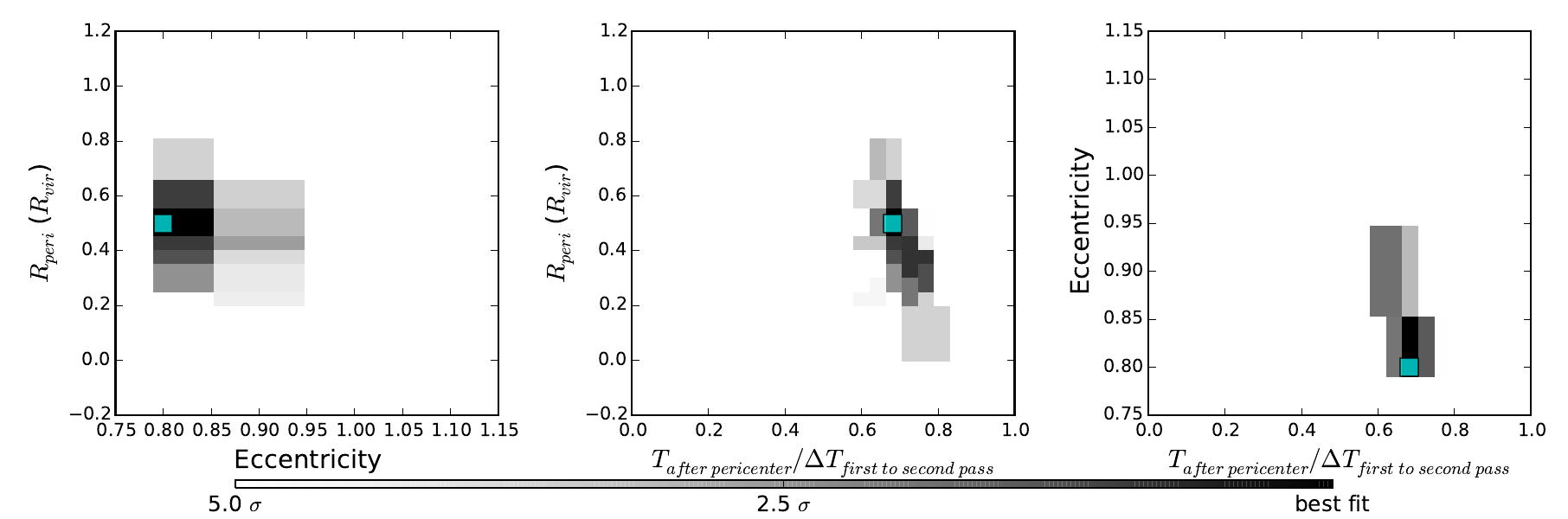}
\caption[Three slices of the score map for UGC 07593]
{Three slices of the score map across pericentric distance, eccentricity, and time for the modeling of UGC 07593. Pericentric distance is in units of the virial radius of the individual galaxy halo, and time is presented as the fraction of time since pericenter over time between the first and second passage. The slices are taken at the best-fit parameter point, shown by the cyan box. the gray scale shows the goodness of the fit from black (the best-fit model) to white (models with scores that are $5\sigma$ or more lower than the score of the best-fit model).}
\label{fig:obssys:ugc07593scoremap}
\end{figure*}


We found a first ever model for UGC 07593. Figure \ref{fig:obssys:ugc07593scoremap} shows slices of the map of scores across three encounter parameters, eccentricity, pericentric distance, and time since pericenter. The slices are taken at the best-fit parameters shown by the cyan box. This system is in a relatively late stage at $0.68\pm^{0.08}_{0.01}$ of the time between the first and the second passages. The reconstructed pericentric distance and eccentricity are $0.50\pm^{0.16}_{0.16}R_{vir}$ and $0.8\pm^{0.05}_{0.05}$, respectively. Taking the physical length and velocity scaling into account we derive the encounter parameters with physical units (See \citealt{Mortazavi:2016hv}). The reconstructed time since the first passage is $27\pm^{30}_{3}$ Myr. The physical pericentric distance is $2.5\pm^{0.6}_{0.7}$ kpc. Both galaxies are almost half-way between a prograde and a polar orbit with the inclinations being at about $45^\circ\pm 20^\circ$ from the orbital plane.


However, the dynamical modeling method we used did not well converge for the other systems. This failure is likely to be the result of different potential factors. In \textbf{UGC 12914} we do not find a smooth rotation in velocity of star-forming components. The gas in the bridge is highly shocked, and powerful HI, CO, radio continuum, and X-ray emission is detected in the bridge, suggesting that the gas in two discs have been stripped off as a result of a virtually head on encounter \citep{Condon:1993ea,Braine:2003cu,2003AJ....126.2171G,Vollmer:2012gr,Appleton:2015ew}. The gas in the bridge does not accompany the rest of the baryons, particularly the collisionless stars, and it can not be reconstructed by \textsc{identikit} test particles. In \textbf{NGC 5257/8} and \textbf{Arp 87} at least one of the galaxies seem to be in polar orbit (\citealt{Privon:2013fs} for NGC 5257/8). According to \cite{Mortazavi:2016hv}, the modeling method we used does not result in a good convergence when any of the galaxies is in a polar or retrograde orbit, so we should have expected similar failure for these two systems.

\section{Shock Fraction vs. Reconstructed Encounter Parameters}
\label{sec:obssys:fshockvsparams}

\begin{table*}
\caption[List of systems with available dynamical models]
{List of systems with available dynamical models. The source
of the model, time since pericenter, time left to coalescence ($\Delta$t),
pericentric separation ($\text{R}_\text{peri}$), and fraction of shocked \ha\ emission ($\text{f}_\text{shocked}$),
are shown. All of these systems except UGC 12914 and Arp 84 have been modeled
with equal mass galaxies. Mass ratio of the models of UGC 12914 and Arp 84 is 3 and 4, respectively .
\cite{Kaufman1999TheCompanion}, \text{\cite{Vollmer:2012gr}} and \text{\cite{Struck:2003kr}} do no provide 
the time left to 
coalescence in their models. The table is sorted by shock fraction.}
\begin{center}
\begin{tabular}{l|l|llll}
system name & source & time (Myr) & $\Delta$t (Myr) & $\text{R}_\text{peri}$ (kpc) & $\text{f}_\text{shocked}$ \\\hline
Arp 84 & \cite{Kaufman1999TheCompanion}&17& - &17.0&0.10\\
NGC 5257/8 &\text{\cite{Privon:2013fs}}  & 230 & 1200 & 21.0 & 0.15 \\
NGC 4676 & \text{\cite{Mortazavi2018HMice}}  & 190 & 775 & 18.0 & 0.26 \\
UGC 12914 &  \text{\cite{Vollmer:2012gr}}  & 26 & - & 1.2 & 0.37 \\
Arp 284 & \text{\cite{Struck:2003kr}} & 170 & - & 6.5 & 0.39 \\
UGC 07593 & this work & 27 & 12 & 2.5 & 0.40 \\
NGC 2623 &  \text{\cite{Privon:2013fs}} & 220 & -80 & 1.0 & 0.90 \\
\end{tabular}
\end{center}
\label{tab:obssys:modeled}
\end{table*}

\begin{figure*}
\begin{center}
\includegraphics[width=0.7\textwidth]{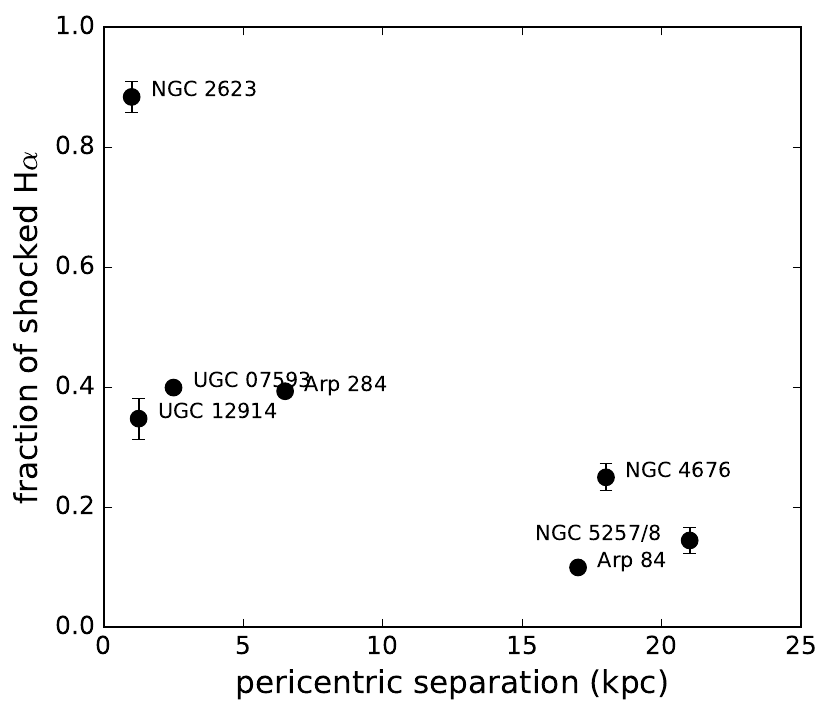}
\caption[Trends of shock fraction in models with available dynamical model]
{Trends of shock fraction in models with available dynamical model. It shows how fraction of shocked \ha\ emission changes with pericentric distance.}
\label{fig:obssys:modeltrend}
\end{center}
\end{figure*}

In this work, we measured the \ha\ kinematics of a sample of 22 galaxy mergers. Nineteen of these mergers are in the stage between the first passage and the coalescence, and the other three are already coalesced. We examine the source of ionization in these mergers using the kinematics and flux ratio of \ha\ and \nii\ emission lines.  We find that the close pairs with projected separation $<25$kpc have a higher \ha\ shock fraction compared to the wide pairs with projected separation $>25$kpc. The average of the shock fraction in the coalesced systems is even higher; shocks are responsible for an average of about 50\% of \ha\ emission in these systems. These results are consistent with results of \cite{Rich:2015kf} suggesting that if the sequence of wide pairs, close pairs, and coalesced mergers is a time sequence, then the shock fraction is growing as the merger proceeds. However, dynamical modeling, which incorporates the complex morphology and kinematics of tidal features, is required to constrain encounter parameters in most galaxy mergers. Determining merger stage only using the separation between the pairs is not alway correct. The geometry of the encounter and observer can result in a close projected separation for a physically wide pair. A pair right after the first passage appears as close as one near coalescence. The impact parameter of the encounter also affects the appearance of projected separation during the encounter. In order to obtain a robust correlation between shock fraction and merger stage we need reconstructed dynamical models of galaxy mergers.

In addition to the model of UGC 07593 from this work, we can find the dynamical models of five other systems in our sample. The Mice galaxies (NGC 4676) are extensively modeled in the literature \citep{Toomre:1972jia,Barnes:2004kp,Privon:2013fs,Mortazavi2018HMice}. NGC 5257/8 and NGC 2623 are among the four systems modeled
in \cite{Privon:2013fs}. We also find a dynamical model of Arp 84, UGC 12914, and Arp 284 in \cite{Kaufman1999TheCompanion}, \cite{Vollmer:2012gr}, and \cite{Struck:2003kr}, respectively. The reconstructed time since the first passage, time left to coalescence, and pericentric separation of these models are shown in Table \ref{tab:obssys:modeled} along with the measured shocked \ha\ fraction from this work.

 Table \ref{tab:obssys:modeled} is sorted by shock fraction. One can see that pericentric separation is  almost sorted in Table \ref{tab:obssys:modeled}. Systems with wide pericentric distance have less fraction of shocked \ha\ than systems with small pericentric distance. This trend can be seen in Figure \ref{fig:obssys:modeltrend}.

\section{Discussion}
 \label{sec:obssys:discussion}

Production of shocks is the result of complex processes and can not be reduced to simple correlations like the ones in Figure \ref{fig:obssys:modeltrend}. However, before coalescence, between the first and second passages, the processes that produce shocks can generally be categorized into two modes \citep{Cox2004GeneratingMergers,Soto:2012hm}. In the tidal mode, the tidal impulse at the time of the first pericenter triggers the shock producing processes, so factors that affect the strength of the tidal force should affect the shock production through this mode. For example, smaller pericentric distance and prograde initial disc orientation should increase the production of shocks via the tidal mode, for they enhance the gravitational tidal effect. In the direct collision mode, the geometry of galactic encounter allows the ISM of the two gas rich discs to directly collide with each other. For this to take place, the two discs should have small enough distance at the pericenter, so that the discs and not just the dark matter halos cross during the first passage. The shocks are produced immediately, and decay as they cool down through radiation. 

\subsection{Shock Production and Encounter Parameters}

Our sample in this work was selected based on the visibility of strong tidal features (see \S \ref{sec:obssys:obs:tgtsel}), so it should be biased toward prograde mergers, and the tidal mode of shock production should be dominant. We predict higher shock fraction in systems that are closer to coalescence, because not only is there more time for the gas inflows to enhance central starbursts/AGNs and produce shocks through outflows, but also violent relaxation, random orbital crossings, induced waves such spiral waves and bars, and the return of tidally striped material during coalescence may generate shocks. In our sample of modeled systems we do not have enough statistics to examine this trend. Notice that in Table \ref{tab:obssys:modeled} we  lack models for time till coalescence for UGC 12914 and Arp 284.

From the spacial distribution of the \ha\ emission, it seems that in UGC 12914 shocks are produced through the direct collision mode. The presence of the continuous gaseous bridge observed on different bands \citep{Condon:1993ea,Braine:2003cu,2003AJ....126.2171G,Vollmer:2012gr,Appleton:2015ew} implies that gas has been striped from the two galaxies because of a recent head-on encounter. According to the model presented by \cite{Vollmer:2012gr}, the relative velocity of galaxies at the time of the encounter was $\approx$ 1000 km/s. The gas clouds in the ISM of the two discs have collided at this speed and dragged themselves away from the stars toward the bridge, If \cite{Vollmer:2012gr} had provided the time until final coalescence, we would expect it to be a long time, for the galaxies are still receding from each other. This would be inconsistent with the trend we expected in the previous paragraph between shock fraction and time left to coalescence, suggesting that the shocks in this galaxy are produced through a fundamentally different procedure.

In both proposed modes of shock production, pericentric separation affects the shock fraction in the same way. The smaller the pericentric separation is, the stronger the induced tidal forces are, and the more shocks via the tidal mode are expected. Similarly, a small enough pericentric separation results in direct collision of gas in the ISM of the two discs, producing shock. Both of these are consistent with the trend seen in Figure \ref{fig:obssys:modeltrend}. We realize that the low statistics of our sample of modeled systems urges us to be caution about this conclusion. On the other hand, we should keep in mind that these scenarios only apply to the time between the first and the second passages. During coalescence, shock production is more complicated and may not be characterized by these simplified modes. In fact, strong shocks may be found in the remnant of a recent coalesced merger, even if it had a large separation at the first passage.

\subsection{The Effect of Mass Ratio}

Assuming that the tidal mode of shock production is dominant in our sample, the shocks should have been created mostly as a result of the mass flows generated by tidal forces during the first passage; consequently, we should find stronger shocks in a galaxy for which the companion is more massive, for tidal force depends on the mass of the perturber. Disregarding other factors, this leads to two predictions about the effect of mass ratio on shock fraction in binary galaxy mergers: First, when galaxies have comparable mass, tidal force on both galaxies should be relatively strong compared to the self-gravity of individual galaxies. In comparison, when one of the companions is much less massive, it may not induce a strong enough tidal force to overcome the self-gravity of its companion and affect the orbit of rotating disc particles, so only one of the companions (the secondary) would experience strong enough tides to create shocks. Therefore, more equal mass mergers are expected to have a higher overall shock fraction. Second, with a similar argument for non-equal mass mergers, we should have seen a higher shock fraction in the smaller companion as it experiences a larger tidal impulse from a more massive perturber. 

The results of \S \ref{sec:obssys:ionization:massratio} confirms the first prediction, but for the second prediction, the evidence points to the contrary. It is important to note that the predictions above would have been valid only if all other factors affecting the shock fraction were the same for all galaxies, and all shocks were created from the tidal mode. However, as discussed earlier, at least for UGC 12914 most of the shocks are not likely to be the consequence of a tidal impulse. Besides, there are other important factors such as gas content and its gravitational stability that may affect the fraction of shocks produced by tidal forces. Without an independent measurement of the gas content of each galaxy we cannot evaluate the effect of these other factors. From a statistical point of view, the scaling relationship between the size, baryonic mass, and gas fraction of nearby spiral galaxies suggest that those with a total baryonic mass $\sim 10^{10.6}\text{M}_\odot$ have the most gas mass reservoir of all star-forming disc galaxies \citep{Wu2018TheGalaxies}. On the other hand, gravitational instability of the gas disc may be a critical player in shock production when the tidal impulse kicks in. According to the Toomre's stability criterion, disc stability against collapse is inversely proportional to gas surface density \citep{Safronov1960OnRotation,Toomre1964OnStars}, and in the nearby Universe, gas surface density is generally maximum for discs with total baryonic mass $\sim 10^{9.7}\text{M}_\odot$ \citep{Wu2018TheGalaxies}. The total baryonic masses of our galaxies are likely to be on either side of these maximum values, so without an accurate constraint on the mass of each galaxy, we cannot evaluate if the effects of gas content explain the low shock fraction of the secondary companions in our minor mergers. Measuring the baryonic mass of the observed galaxies is out of the scope of this work.

It is important to emphasize that in this work, all of the relationships between shock fraction and encounter parameters are based on either a small sample of 22 observed systems or an even smaller sample of 7 systems with available dynamical models. Besides, taking the extinction into account may significantly lower the measurement of shock fraction or increase its uncertainty. In order to independently investigate the effect of each parameter and possibly find more complicated relations, we need a larger statistical sample of galaxy mergers observed with optical IFU instruments. Such observations could be used to understand the possible role of merger-induced shocks in the overall quenching of star formation in the Universe. 
Ongoing and future IFU galaxy surveys (e.g. SAMI: \citealt{Croom:2012fo}, MaNGA: \citealt{Bundy:2015ft}, etc.) provide a promising area for further investigation in near future. 

\section*{Summary}

In this work we observed 22 galaxy mergers with strong tidal features, using the SparsePak IFU on the WIYN telescope at KPNO. We reduce the data and analyze the emission lines with one- and two-component Gaussians. We use an MCMC code to estimate the uncertainty of fit parameters, and select the best number of components using the F-test. Relatively high spectral resolution of our data allows us to use velocity dispersion of emission lines along with \nii/\ha\ line ratios to separate \ha\ originating at star-forming regions from that arising from shocked gas. We use emission line maps to confirm that most of high velocity dispersion and high \nii/\ha\ components are galaxy wide shocks, likely to be induced as a result of interaction. We found that the fraction of \ha\ emission from shocked gas is correlated with the separation of galaxies in pairs. Close pairs have higher shock fraction than wide pairs. The three coalesced systems show the highest average shocked \ha\ fraction.

We use the modeling method developed in \cite{Mortazavi:2016hv} along with the velocity maps of star-forming regions to model the dynamics of equal mass pairs in our sample. We find the first ever constraints on the encounter parameters of UGC 07593, but obtain poor convergence in the other four attempts. We find dynamical model of five other systems in our sample from the literature, and explore the correlations between some encounter parameters and shock fraction. In these systems pericentric separation appear to be correlated with the fraction of shocked \ha. We suggest two modes of shock production that are responsible for most of the shocks in the early stages of a merger, after the first passage and before the coalescence.

\section*{Acknowledgement}
The authors would like to thank the referees for their insightful suggestions. We also thank Colin Norman, Joshua Barnes, David Law, Gregory Snyder, Susan Kassin, Ron Allen, and Luciana Bianchi for valuable discussions throughout this work. This project was supported in part by the Space Telescope Science Institute (STScI) Director's Discretionary Research Fund (DDRF). This work used the Extreme Science and Engineering Discovery Environment (XSEDE), which is supported by National Science Foundation grant number ACI-1053575 (see \citealt{Towns2014XSEDE:Discovery}). We used the DSS scan of Arp 273 in this work. The DSS was produced at the STScI under U.S. Government grant NAG W-2166. The images of these surveys are based on photographic data obtained using the Oschin Schmidt Telescope on Palomar Mountain and the UK Schmidt Telescope. The plates were processed into the present compressed digital form with the permission of these institutions.








\bibliographystyle{mnras}
\bibliography{Mendeley} 




\appendix

\section{Notes on Individual Systems}
\label{sec:individual}
In the Appendix we discuss some of the interesting features we found in two
of the systems observed in this work.

\begin{figure*}
\begin{center}
\subfloat[]{\label{fig:appendix:arp284group1}\includegraphics[height=0.33\textwidth]{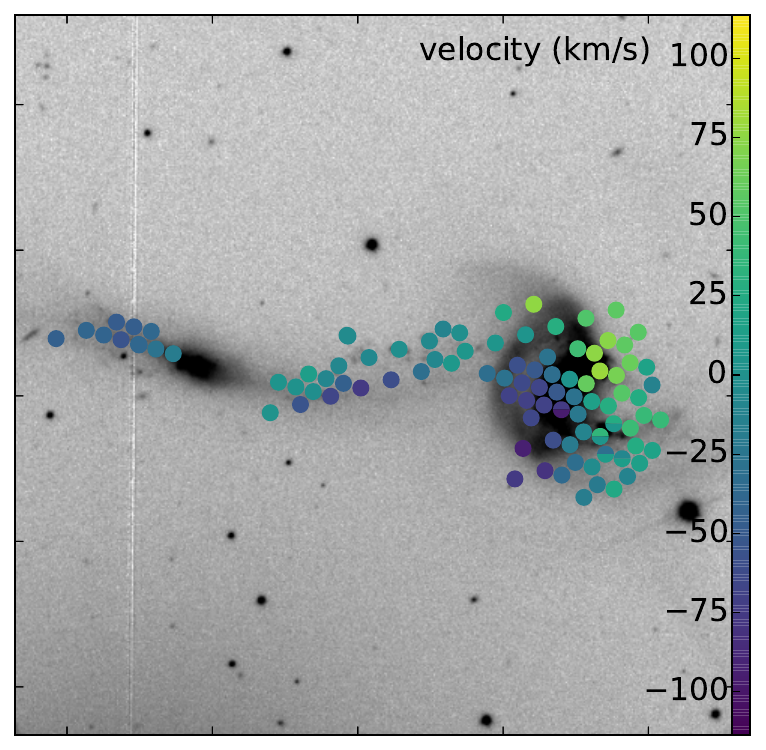}}
\subfloat[]{\label{fig:appendix:arp284group2_3}\includegraphics[height=0.33\textwidth]{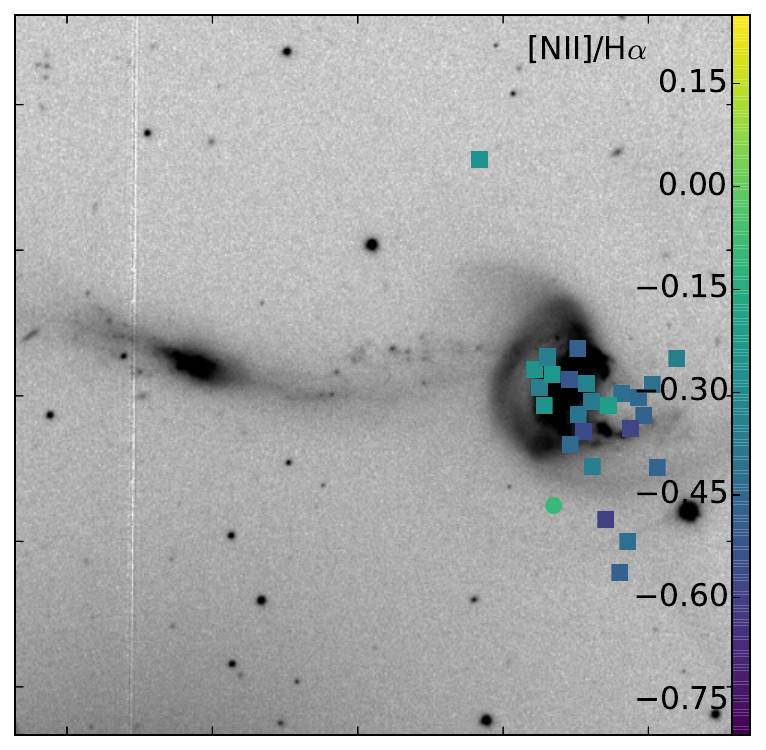}}
\subfloat[]{\label{fig:appendix:arp284diff}\includegraphics[height=0.33\textwidth]{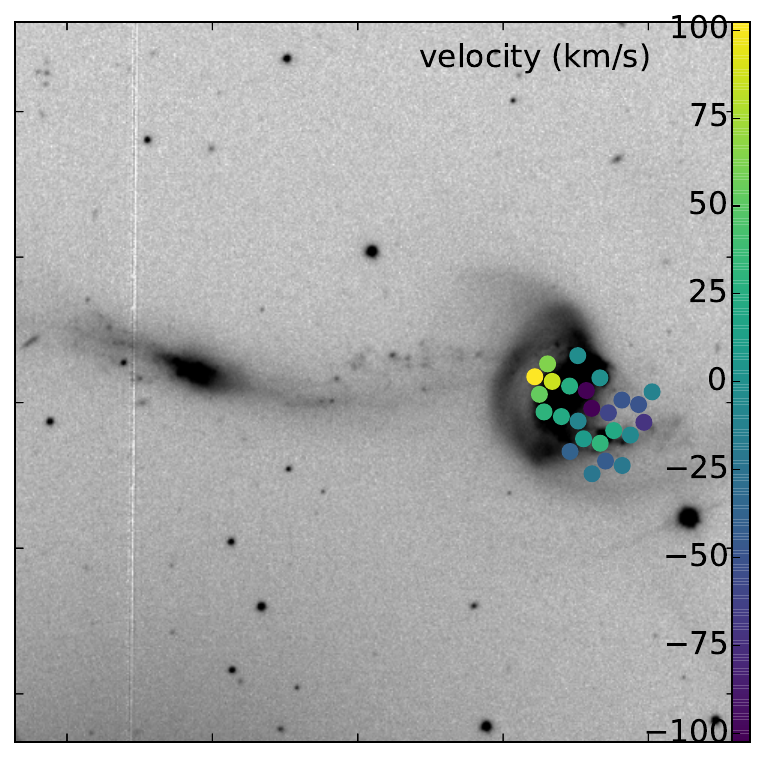}}
\caption[\ha\ maps of Arp 284]
{\textbf{Arp 284:} (a) Velocity map of star-forming components. In the secondary galaxy (east) there
is no \ha\ emission near the center. Dynamical modeling of \protect\cite{Struck:2003kr} suggests 
that the secondary has been loosing gas to the primary since the first passage about 170 Myr 
ago, leaving it quiescent, particularly on the western side. (b) Map of \nii/\ha\ for components in Groups 2 and 3 likely originated at shocked gas. They show a cone-like structure, which is generally indicative of outflows. (Group 2 is shown with circles and Group 3 with squares) (c) Map of the difference in the velocity of broad and narrow
components in fibers where two-component fit is preferred. Blue indicates that the broad component is blueshifted, i.e. it is likely to be outflowing gas, and yellow indicates that the broad component is redshifted, indicating that is likely to be inflowing gas. On the right (west), which is the wide end of the cone the circles are blue, meaning that the broad 
components are blueshifted. This is consistent with outflows. However, On the left (east), on narrow side of the cone and close to the center of the primary galaxy, the yellow circles indicate that the broad components are redshifted. This suggests that the broad components are the inflowing gas striped from the secondary galaxy, which is consistent with the dynamical model of \cite{Struck:2003kr}.}
\end{center}
\end{figure*}

\textbf{Arp 284:}
In the secondary galaxy (NGC 7715, east) we do not detect \ha\ emission in the western side of the edge-on disc which is toward the primary galaxy. This can be seen is shown in Figure \ref{fig:appendix:arp284group1}. There are HII regions in the tidal bridge between the galaxies. \cite{Smith:1992bo} found an 
HI bridge in the same position as the optical bridge. The primary galaxy (NGC 7714, west) displays smooth rotation in the star-forming 
components, which is scattered throughout the tilted disc.

This system is studied extensively in the literature. \cite{Delgado:1998bf}  used
HST/GHRS ultraviolet spectroscopy and ground based radio, optical, and X-ray observations to perform a spectral synthesis modeling on central
300 pc of the primary galaxy. They suggest that the center of this galaxy contains a very young starburst (4.5 Myrs) along with an older stellar population with ages around tens of Myrs or older. Dynamical modeling by \cite{Struck:2003kr} indicates that a significant amount of gas has been transferring from the secondary companion to the primary galaxy since the first passage, which 
occurred $\approx 170$ Myr ago. This gas transfer is fueling the starburst in the core of NGC 7714. It seems that tidal striping of gas and lack of inflows is why the secondary galaxy has quenched star formation. 
\cite{Smith:2005he} observed X-ray emission in
the core of the primary galaxy attributing it to mechanical energy injected into to the ISM
by either supernovae or High Mass 
X-ray Binaries (HMXBs),
both as results of the central starburst.
In addition, Infrared spectroscopy with 
Spitzer space telescope shows
no evidence of an obscured AGN 
in the center of the primary, confirming that it is a young unobscured starburst galaxy \citep{Brandl:2004ew}.

In the primary galaxy, we find a vertical cone-like structure made mostly by Group 3 and some Group 2 components, which is shown in Figure \ref{fig:appendix:arp284group2_3}. At first glance, we speculate that this structure and its low overall \nii/\ha\ ratio indicates
that they are the outflows from the central starburst proposed by \cite{Smith:2005he}.
However, by plotting the difference in the velocity of broad and narrow components in fibers where two-component fit is preferred, we come up with a new conclusion. Figure \ref{fig:appendix:arp284diff} shows the velocity of narrow components subtracted from the velocity of broad components, so blue (yellow) circles indicate that the broad component is blueshifted (redshifted) with respect to the narrow component. Assuming that the narrow component corresponds to the star-forming region in the disc and that we only see the broad component that is physically in front of the narrow component, we can argue that blueshifted broad components show outflows and redshifted ones show inflows. In Figure \ref{fig:appendix:arp284diff} we see blueshifted broad components on the wide end of the cone confirming our speculation of them showing outflows. However, on the narrow end of the cone closer to the center, we see redshifted broad components, which can be explained by the inflowing gas. This could be the gas that is striped from the secondary galaxy leaving it quiescent on its west side. This observation is consistent with the dynamical model of \cite{Struck:2003kr} which suggested that the secondary galaxy is depleting gas into the center of the primary one, causing a starburst and generating the outflows toward the west. This observation demonstrates that by properly separating kinematic components in high spectral resolution nebular emission data, not only outflows but also inflows may be detected.

\begin{figure*}
\subfloat[]{\label{fig:appendix:ugc12914group1}\includegraphics[height=0.33\textwidth]{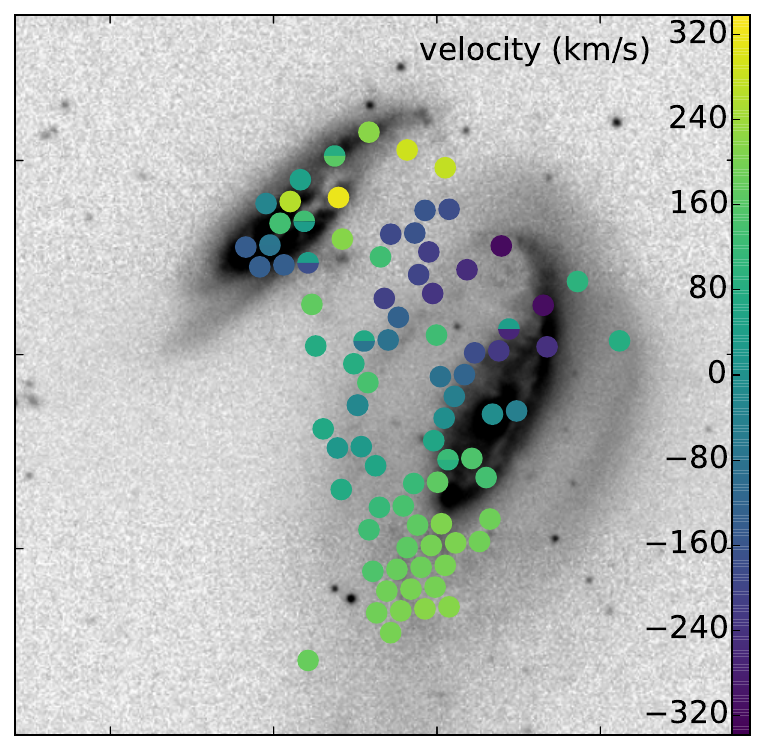}}
\subfloat[]{\label{fig:appendix:ugc12914group2_3}\includegraphics[height=0.33\textwidth]{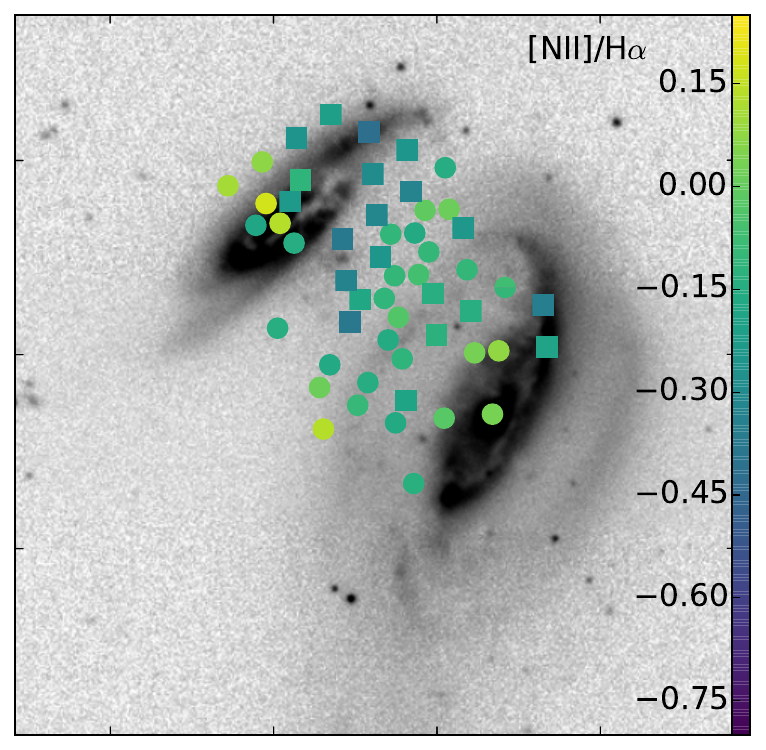}}
\subfloat[]{\label{fig:appendix:ugc12914diff}\includegraphics[height=0.33\textwidth]{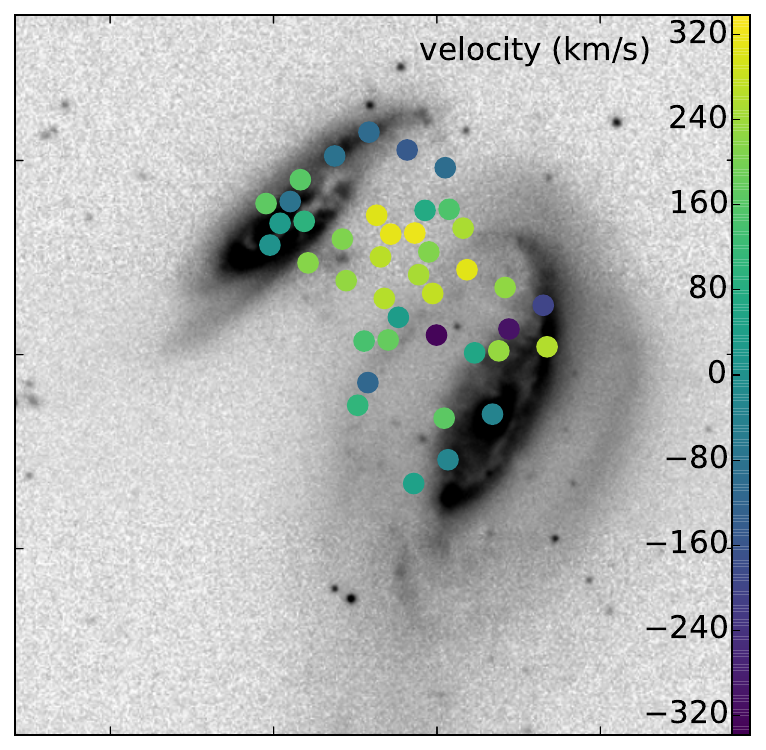}}
\caption[\ha\ maps of UGC 12914/5]
{\textbf{UGC 12914/5:} (a) Velocity map of star-forming regions. The strong \ha\ emission originating from the bridge, which is
relatively faint in the SDSS r-band image, is consistent with the strong HI, CO, radio continuum, and X-ray emissions observed in the bridge in previous works. \protect\cite{Vollmer:2012gr} proposes a dynamical model in which the galaxies have had 
a head-on collision about 26 Myr ago. The collision has striped off the gas from both discs into the bridge between them. (b) Map of \nii/\ha\ ratio for components in Groups 2 and 3 that are likely to originate from shocked gas. They are located mostly in the bridge region, suggesting that the shocks are generated as a result of direct collision of gas in the ISM of the two discs. (c) Map of the difference between the velocities of the broad and narrow components in fibers where the two-component fit is preferred by the F-test. The velocity difference $\approx$ 300 km/s is consistent with the double peaks observed in HI line profiles \cite{Condon:1993ea} and the extended X-ray emission in the bridge which are probably originating from shocks of similar 
velocities \protect\citep{Appleton:2015ew}.}
\label{fig:appendix:ugc12914maps}
\end{figure*}

\textbf{UGC 12914/5:}
This system is also known as the ``Taffy galaxies", because HI and 4.86 GHz observations by \cite{Condon:1993ea} showed a joined 
HI and radio continuum bridge between the two galaxies. \cite{Condon:1993ea} suggest that cosmic rays, magnetic fields, and HI gas
have been striped from the two galaxies as the result of a nearly head-on collision, about 20 Myrs ago. In the bridge, they found double
peaks in the HI profile separated by 100-300 km/s. \cite{Braine:2003cu} finds a significant amount of molecular gas emission (CO) in the bridge, suggesting that 18-35\% of the total gas mass in the system sits in the bridge. The CO emission does not reveal the double peak
and only the high-velocity HI component has a CO counterpart.

\cite{Vollmer:2012gr} modeled the dynamics of this system using a model that includes collisionless halo+stars particles
and collisional sticky gas particles. They distinguish the molecular gas from neutral hydrogen using a prescription for total gas density. They found the best-fit model at about
26 Myr after the first pass. Their model reproduced the morphology and kinematics of stars and gas in this system including the prominent gaseous bridge. They also reproduced 
the double-component HI profile and the single component CO profile in the bridge region. Changing the cloud-cloud collision parameter in their simulation, they argued that the double component profile is produced mainly by the 
distortion caused by the collisional nature of ISM, and not by the tidal distortions.

\cite{Appleton:2015ew} used the Chandra observatory to show that the bridge also emits a significant amount of X-ray. They showed that 19\% of the X-ray luminosity of the system comes from the bridge. They also used Herschel Far-IR data to estimate the SFR in the bridge, concluding that SFR is too low to account for X-ray emission via outflows. Moreover, they showed
that the peak of the diffuse X-ray emission does not match with the peak of the radio continuum, ruling out a direct connection between the X-ray and synchrotron emission caused by cosmic
rays in the bridge. They suggest that the main source of X-ray emission in the bridge is shock heating due to collision of the ISM in the 
two galaxies as was suggested earlier by \cite{Struck:1997km}. The shocked gas can be heated up to $\sim 10^6$ K during the collision and cool down to $\sim 10^5$ K in 35 Myr and to $\sim 10^4$ K within less than 100 Myr. Based on their estimated time of $\sim 26$ Myr after the pericenter, the temperature of the observed soft X-ray is consistent with shocks with speed range of 430-570 km/s.

Figure \ref{fig:appendix:ugc12914diff} shows the velocity difference between the broad and narrow component in fibers where two component fit is preferred by the F-test. It shows that most
of the fibers in the bridge region prefer a two-component fit. In most of these fibers, the narrow component meets the criteria 
for a normal star-forming region, and the broad component is considered as shocked. In most of these two-component fits the broad component is redshifted with respect to the narrow component ($>300$ km/s), suggesting that outflows caused by star formation are not responsible for producing the shocks in this region. This velocity difference is consistent with the velocity difference between peaks in the HI profile in the bridge. Based on the velocity difference we can argue that the narrow
star-forming components shown in Figure \ref{fig:appendix:ugc12914group1} correspond to the HI components with corresponding CO lines, and the broad shocked components shown in Figure \ref{fig:appendix:ugc12914group2_3}
correspond to the HI components without corresponding CO lines. The velocity difference is a bit lower but consistent with shock speeds required to produce the X-ray emission in the bridge region. Therefore, our data confirms
that the shocked components in the bridge are produced as a result of the fast direct collision of gas in the ISM at the time of the first passage.

In Figure \ref{fig:appendix:ugc12914group1} we also find a void of star forming components in the center of UGC 12914. \cite{Appleton:2015ew}
found a slightly extended ultra-luminous X-ray source in this location, hinting to the presence of an obscured AGN. Though, the velocity dispersion of the shocked components are not
as high as what is expected for AGN hosts \citep{Rich:2014ib}.


\bsp	
\label{lastpage}
\end{document}